\documentclass[10pt,twocolumn,twoside]{IEEEtran}
\usepackage{ifpdf}
\usepackage{cite}
\usepackage[dvips]{graphicx}
\usepackage[cmex10]{amsmath}
\usepackage{amssymb}
\usepackage{amsmath}
\usepackage{mathrsfs}
\usepackage[lined,boxed,commentsnumbered, ruled]{algorithm2e}
\usepackage{array}
\usepackage{fmtcount}
\usepackage{mdwmath}
\usepackage{eqparbox}
\usepackage[tight,footnotesize]{subfigure}

\newtheorem{lemma}{{Lemma}}
\newtheorem{theorem}{{Theorem}}

\newtheorem{proposition}{{Proposition}}

\usepackage[caption=false,font=footnotesize]{subfig}
\newcommand{\bs}{\boldsymbol}
\usepackage{url}

\begin{document}

\author{Yuanming~Shi,~\IEEEmembership{Student Member,~IEEE,}
        Jun~Zhang,~\IEEEmembership{Member,~IEEE,}
        and~Khaled~B. Letaief,~\IEEEmembership{Fellow,~IEEE}
\thanks{The authors are with the Department of Electronic and Computer Engineering,
Hong Kong University of Science and Technology (e-mail: \{yshiac, eejzhang,
eekhaled\}@ust.hk).}}
\title{Optimal Stochastic Coordinated Beamforming for Wireless Cooperative Networks with CSI Uncertainty}
\maketitle

\begin{abstract}
Transmit optimization and resource allocation for wireless cooperative networks with channel state information (CSI) uncertainty are important but challenging problems in terms of both the uncertainty modeling  and  performance optimization. In this paper, we establish a generic stochastic coordinated beamforming  (SCB) framework that provides flexibility in  the channel uncertainty modeling, while guaranteeing optimality in the transmission strategies. We adopt a general stochastic model for the CSI uncertainty, which is applicable for various practical scenarios. The SCB problem turns out to be a joint chance constrained program (JCCP) and is known to be highly intractable. In contrast to all the previous algorithms for JCCP that can only find \emph{feasible but sub-optimal} solutions, we propose a novel stochastic DC (difference-of-convex) programming algorithm with \emph{optimality guarantee}, which can serve as the benchmark for evaluating heuristic and sub-optimal algorithms. The key observation is that the highly intractable probability constraint can be equivalently reformulated as a DC constraint. This further enables efficient algorithms to achieve optimality. Simulation results will illustrate the convergence, conservativeness, stability
and performance gains of the proposed algorithm.     
\end{abstract}

\section{Introduction}
\IEEEPARstart{N}{etwork} cooperation is a promising way to improve both energy efficiency and spectral efficiency of wireless networks by sharing control information and/or user data\cite{Gesbert_JSAC10}. Among all the cooperation strategies, jointly processing the user data can achieve the best performance by exploiting the benefits of a large-scale virtual MIMO system \cite{Foschini_2006network,Jun_2009networked}. This inspires a recent proposal of a new network architecture, i.e., Cloud radio access network (Cloud-RAN)\cite{mobile2011c,Yuanming_TWC2014}, which will enable fully cooperative transmission/reception by moving all the baseband signal processing to a datacenter Cloud.  In order to fully exploit the benefits of cooperative networks and develop efficient transmission strategies (i.e., coordinated beamforming), channel state information (CSI) is often required. However, in practical scenarios, inevitably there will be uncertainty in the obtained channel coefficients, which may originate from a variety of sources. For instance, in frequency-division duplex (FDD) systems, the CSI uncertainty may originate from downlink training based channel estimation \cite{Jindal_TC2010unified} and uplink limited feedback \cite{love2008overview}. It could also be due to the hardware deficiencies, delays in CSI acquisition \cite{Tse_TIT2012completely,Jun_2009mode}, and partial CSI acquisition \cite{Yuaning_ICC2014,Luo_arXiv2013}. With full and perfect CSI, efficient performance optimization can often be achieved through convex formulations, e.g., coordinated beamforming via second-order cone programming \cite{WeiYu_WC10,Yuanming_TWC2014}. However, the channel knowledge uncertainty due to the partial and imperfect CSI brings technical challenges in system performance optimization. 

To address  such challenges brought by the channel knowledge uncertainty, one may either adopt a robust optimization formulation \cite{ben2009robust} or stochastic optimization formulation \cite{shapiro2009lectures}. Specifically, for the robust formulation, the channel knowledge uncertainty model is deterministic and set-based \cite{Emil_TSP2012}. Thus, the corresponding transmission strategies aim at guaranteeing the worst-case performance over the entire uncertainty set. The primary advantage of robust formulation is the computational tractability \cite{bertsimas2011theory}. However, the worst-case formulation might be over-conservative \cite{bertsimas2011theory}, as the probability
of the worst case could be very small \cite{gershman2010convex}. Meanwhile, how to model the uncertainty set is also challenging \cite{Bjornson_TCIT2013}. On the other hand, in the stochastic optimization formulation, the channel knowledge is modeled by a probabilistic description. Thus, the corresponding transmission strategies seek to immunize a solution against the stochastic uncertainty in a probabilistic sense \cite{Angela_TSP10,Angela_JSAC2013,Vicent_TSP2013,Xiaodong_2014power,wang2011probabilistic}. The freedom of the probabilistic
robustness can provide improved system performance \cite{Xiaodong_2014power} and provide a tradeoff between the conservativeness and probability guarantee \cite{bertsimas2011theory}.

Motivated by the fact that most wireless systems can tolerate occasional outages
in the quality-of-service (QoS) requirements\cite{Angela_TSP10,Angela_JSAC2013,Vicent_TSP2013}, in this paper, we propose a stochastic coordinated beamforming (SCB) framework to minimize the total transmit power while guaranteeing the system probabilistic QoS requirements. In this framework, we only assume that the distribution information of the channel uncertainty is available, but without any further structural modeling assumptions (e.g., adopting the ellipsoidal error model for robust design \cite{Emil_TSP2012} or assuming complex Gaussian random distribution for the channel errors \cite{Vicent_TSP2013,Xiaodong_2014power,wang2011probabilistic} for stochastic design). In spite of the distinct advantages, including the design flexibility and the insights
obtained by applying the SCB framework to handle the CSI uncertainty, it falls into a joint chance constrained program (JCCP) \cite{shapiro2009lectures},
which is known to be highly intractable \cite{hong2011sequential}. All the available algorithms (e.g., the scenario approach \cite{Campi_SIAM2008,Angela_JSAC2013,Yuaning_ICC2014} and the Bernstein approximation method \cite{nemirovski2006convex, Angela_TSP10,Vicent_TSP2013, wang2011probabilistic}) can only find \emph{feasible but suboptimal} solutions without any optimality guarantee. 

In contrast, in this paper, we propose a novel stochastic DC programming algorithm, which can find the globally optimal solution if the original SCB problem is convex and find a locally optimal solution if the problem is non-convex. The main idea of the algorithm is to reformulate the system probabilistic QoS constraint as a DC constraint, producing an equivalent stochastic DC program. 
Although the DC programming problem is still non-convex, it has the algorithmic advantage and can be efficiently solved by the successive convex approximation algorithm  \cite{Luo2013unified, hong2011sequential}.  

The main computational complexity of the proposed algorithm comes from solving a  large-sized sample
problem with the Monte Carlo approach at each iteration. This makes such an approach inapplicable in large-size networks.  However,  the proposed stochastic DC programming algorithm  gives  a first attempt to solve a highly-intractable and highly-complicated problem with optimality guarantee, while existing algorithms fail to possess the optimality feature. Therefore, it can serve as a performance benchmark for evaluating other suboptimal and heuristic algorithms.

\subsection{Related Works}
The chance constrained programming has recently received emerging interests in designing efficient resource allocation strategies in communication networks by leveraging the distribution information of uncertain channel knowledge \cite{Angela_TSP10,Angela_JSAC2013,Vicent_TSP2013,Xiaodong_2014power,wang2011probabilistic,Yuaning_ICC2014,Luo_arXiv2013}. However, due to the high intractability of the underlying chance or probabilistic constraints  (e.g., it is difficult to justify the convexity or provide analytical expressions), even finding a feasible solution is challenging. Therefore, it is common to approximate the probability constraint to yield  computationally tractable and deterministic formulations. One way is to approximate the chance constraints using analytical functions, which, however, often requires further assumptions on the distribution of the uncertain channel knowledge (e.g., complex Gaussian distributions for Bernstein-type inequality approximation \cite{Vicent_TSP2013,wang2011probabilistic} or the affine constraint functions in perturbations for Bernstein approximation \cite{nemirovski2006convex,Xiaodong_2014power,Angela_TSP10}). The other way is to use the Monte Carlo simulation approach to approximate the chance constraints (e.g., the scenario approach \cite{Campi_SIAM2008,Angela_JSAC2013,Yuaning_ICC2014} and the conditional-value-at-risk (CVaR) \cite{rockafellar2000optimization}). However, all the above approaches only seek conservative approximations to the original problem. Thus, it is difficult to prove the optimality and quantify the conservativeness of the obtained solutions.

Hong {\emph{et. al}} \cite{hong2011sequential} recently made a breakthrough on providing optimality of the highly intractable joint chance constrained programming problems for the first time. However, the convexity of the functions in the chance constraint is required. Our proposed stochastic DC programming algorithm is inspired by the ideas in \cite{hong2011sequential}. Unfortunately, the  functions in the chance constraint in our problem are non-convex, and thus, we cannot directly apply the algorithm in \cite{hong2011sequential}. Instead, by exploiting the special structure of the functions in the chance constraint, we equivalently reformulate the chance constraint into a DC constraint. The resulting DC program is further supported by efficient algorithms. Thus, we extend the work \cite{hong2011sequential} by removing the convexity assumption on the functions in the chance constraint. Furthermore, to improve the convergence rate, instead of fixing the approximation parameter as in \cite{hong2011sequential}, a joint approximation method is proposed.

\subsection{Contributions}
In this paper, we provide a general framework to design optimal transmission strategies with CSI uncertainty for wireless cooperative networks. The major contributions are summarized as follows:
\begin{enumerate}
   
\item We establish a general SCB framework to
cope with the uncertainty in the available channel knowledge, which intends to minimize the total transmit power with a system probabilistic QoS guarantee. This framework only requires the distribution information of the  uncertain channel coefficients. Thus, it enjoys the flexibility in modeling channel knowledge uncertainty without any further structural assumptions. The SCB problem is then formulated as a JCCP problem.

\item We develop a novel stochastic DC programming algorithm  to solve the
SCB problem,  which will converge to the globally optimal solution if the SCB problem is convex or a locally optimal solution if it is non-convex. The proposed stochastic DC programming algorithm can be regarded as the first attempt to guarantee the optimality for the solutions of JCCP without the convexity assumption on functions in the chance constraint \cite{hong2011sequential}, while the available algorithms (i.e., the scenario approach and the Bernstein approximation method) for JCCP can only find a feasible solution without any optimality guarantee.  
\item The proposed SCB framework is simulated in Section {\ref{sim}}. In particular, the convergence, conservativeness, stability and performance gains of the proposed algorithm are illustrated.   
\end{enumerate}

\subsection{Organization}
The remainder of the paper is organized as follows. Section II presents the
system model and problem formulation, followed by the problem analysis. In Section {\ref{DCA}}, the stochastic DC programming algorithm is developed.  Simulation results will be presented in Section {\ref{sim}}. Finally, conclusions
and discussions are presented in Section {\ref{con_dis}}. To keep the main
text clean and free of technical details, we divert most of the proofs to
the appendix.

\section{System Model and Problem Formulation}
We consider a fully cooperative network{\footnote{ The proposed framework can be easily extended to more general cooperation scenarios as shown in \cite{Emil_TSP2012}.}} with $L$ radio access units (RAUs), where the $l$-th RAU is equipped with $N_{l}$ antennas, and there are $K$ single-antenna mobile
users (MUs). The centralized signal processing is performed at a central processor, e.g., at the baseband unit (BBU) pool in Cloud-RAN \cite{Yuanming_TWC2014}. The propagation channel from the $l$-th RAU to the $k$-th MU is denoted as
${\bf{h}}_{kl}\in\mathbb{C}^{N_{l}}, 1\le k\le K,1\le l\le L$. We focus on
the downlink transmission, for which the joint signal  processing is more
challenging. The received signal $y_{k}\in\mathbb{C}$ at MU $k$ is given by
\setlength\arraycolsep{1pt}
\begin{eqnarray}
y_{k}=\sum_{l=1}^{L}{\bf{h}}_{kl}^{\sf{H}}{\bf{v}}_{lk}s_{k}+\sum_{i\ne k}\sum_{l=1}^{L}{\bf{h}}_{kl}^{\sf{H}}{\bf{v}}_{li}s_{i}+n_{k},
\forall k,
\end{eqnarray}
where $s_{k}$ is the encoded information symbol for MU $k$ with $\mathbb{E}[|s_{k}|^2]=1$,
${\bf{v}}_{lk}\in\mathbb{C}^{N_{l}}$ is the transmit beamforming vector from
the $l$-th RAU to the $k$-th MU, and $n_{k}\sim\mathcal{CN}(0, \sigma_{k}^2)$
is the additive Gaussian noise at MU $k$. We assume that $s_{k}$'s and $n_{k}$'s
are mutually independent and all the users apply single user detection. The
corresponding signal-to-interference-plus-noise ratio (SINR) for MU $k$ is
given by
\begin{eqnarray}
\label{snrr}
{\Gamma}_{k}({\bf{v}}, {\bf{h}}_{k})={{|{\bf{h}}_{k}^{\sf{H}}{\bf{v}}_{k}|^2}\over{\sum_{i\ne
k}|{\bf{h}}_{k}^{\sf{H}}{\bf{v}}_{i}|^2+\sigma_{k}^2}}, \forall k,
\end{eqnarray}
where ${\bf{h}}_{k}\triangleq [{\bf{h}}_{k1}^{T},{\bf{h}}_{k2}^{T},\dots,{\bf{h}}_{kL}^{T}]^{T}=[h_{kn}]_{1\le
n\le N}\in\mathbb{C}^{N}$ with $N=\sum_{l=1}^{L}N_{l}$, ${\bf{v}}_{k}\triangleq[{\bf{v}}_{1k}^{T},
{\bf{v}}_{2k}^{T},\dots, {\bf{v}}_{Lk}^{T}]^{T}\in\mathbb{C}^{N}$ and ${\bf{v}}\triangleq
[{\bf{v}}_k]_{k=1}^K\in\mathbb{C}^{NK}$. The beamforming vectors ${\bf{v}}_{lk}$'s
are designed to minimize the total transmit power while satisfying the QoS
requirements for all the MUs.  The beamformer design problem can be formulated
as
\begin{eqnarray}
\mathscr{P}_{\textrm{Full}}:
\mathop {\rm{minimize}}_{{\bf{v}}\in\mathcal{V}}&&\sum_{l=1}^{L}\sum_{k=1}^{K}\|{\bf{v}}_{lk}\|^2\nonumber\\
{\rm{subject~to}}&&\Gamma_{k}({\bf{v}}, {\bf{h}}_{k})\ge \gamma_{k},
\forall k, 
\end{eqnarray} 
where $\gamma_{k}$ is the target SINR for MU $k$, and the convex set $\mathcal{V}$
is the feasible set of ${\bf{v}}_{lk}$'s that satisfy
the per-RAU power constraints:
\begin{eqnarray}
{\mathcal{V}}\triangleq\left\{{\bf{v}}_{lk}\in\mathbb{C}^{N_{l}}: \sum\limits_{k=1}^{K}\|{\bf{v}}_{lk}\|^2\le
P_{l}, \forall l, k\right\},
\end{eqnarray}
with $P_{l}$ as the maximum transmit power of the RAU $l$. 

The problem $\mathscr{P}_{\textrm{Full}}$ can be reformulated as a second-order conic programming (SOCP) problem, which is convex and can be solved
efficiently (e.g., via the interior-point method). Please refer to \cite{Yuanming_TWC2014}
for details. Such coordinated beamforming can significantly improve the network
energy efficiency. However, solving problem $\mathscr{P}_{\textrm{Full}}$
requires full and perfect CSI available at the central processor. In practice, inevitably there will be uncertainty in the available channel knowledge. Such uncertainty may originate from various sources, e.g., training based channel estimation \cite{Jindal_TC2010unified}, limited feedback \cite{love2008overview}, delays \cite{Tse_TIT2012completely, Jun_2009mode}, hardware deficiencies \cite{Emil_TSP2012} and partial CSI acquisition \cite{Yuaning_ICC2014,Luo_arXiv2013}. In the next subsection, we will provide a generic stochastic model for the CSI uncertainty. 

\subsection{Stochastic Modeling of CSI Uncertainty}
In this paper, we only assume that the distribution information  of the channel knowledge ${\bf{h}}=[{\bf{h}}_{k}]_{k=1}^{K}\in\mathbb{C}^{NK}$ is available. That is, ${\bf{h}}$ is a random vector drawn from the support set $\Xi\in\mathbb{C}^{NK}$ with the distribution as $\mathbb{P}$. This helps  avoid any structural assumptions on the deterministic  channel uncertainty models and the assumptions on the distribution types of the stochastic channel uncertainty models. In the following, we will provide three examples to justify such a stochastic model. 

\subsubsection{Example  One (Additive Error Model)}  
The following additive error model is commonly used to model the uncertainty of CSI acquisition
\begin{eqnarray}
{\bf{h}}_{k}=\hat{\bf{h}}_k+{\bf{e}}_k, \forall k,
\end{eqnarray} 
where $\hat{\bf{h}}_k$'s are the estimated imperfect channel coefficients and ${\bf{e}}_k$'s are the estimation error vectors.  To facilitate the Bernstein-type inequality approximation for the chance constrained programming, one may assumes that the error vectors follow the complex Gaussian distribution \cite{Xiaodong_2014power,Vicent_TSP2013,wang2011probabilistic}, i.e.,
${\bf{e}}_k\in\mathcal{CN}({\bf{0}}, {\bs{\Theta}}_k), \forall k$,
where ${\bf{\Theta}}_k\in\mathbb{H}^{N}$ with ${\bs{\Theta}}_k\succeq{\bf{0}}$,  is the covariance matrix of the error vector ${\bf{e}}_k$. Based on this model, we can reconstruct the distribution of the channels as
\begin{eqnarray}
{\bf{h}}_k\sim\mathcal{CN}(\hat{\bf{h}}_k, {\bs{\Theta}}_k), \forall k.
\end{eqnarray} 
To leverage the robust design, one often
further assumes the following ellipsoidal channel uncertainty model to bound the
errors in deterministic sets, i.e., ${\bf{e}}_{k}\tilde{\bf{\Theta}}_k{\bf{e}}_{k}^{\sf{H}}\le 1, \forall k$, where $\tilde{\bf{\Theta}}_k\in\mathbb{H}^{N}$ with $\tilde{\bf{\Theta}}_k\succeq{\bf{0}}$
specifies the shape and size of the ellipsoid of the error vector ${\bf{e}}_k$ \cite{Emil_TSP2012}.

\subsubsection{Example Two (Gauss-Markov Uncertainty Model)}
\label{gauss}
The imperfect CSI can also be modeled as the following Gauss-Markov model \cite{Jeffrey_2011mimo}:
\begin{eqnarray}
\label{csierror}
{\bf{h}}_{kl}={\bf{R}}_{kl}^{1/2}\underbrace{(\sqrt{1-\tau_{kl}}\hat{\bf{c}}_{kl}+\tau_{kl}{\bf{e}}_{kl})}_{{\bf{c}}_{kl}},
\forall k, l,
\end{eqnarray}
where ${\bf{R}}_{kl}\in\mathbb{H}^{N_{l}\times N_{l}}$ with ${\bf{R}}_{kl}\succeq {\bf{0}}$ is the channel
correlation matrix between MU $k$ and RAU $l$, $\hat{\bf{c}}_{kl}\in\mathcal{CN}({\bf{0}},{\bf{I}}_{N_l})$ is the
imperfect estimate of the true channel vector ${\bf{c}}_{kl}$ and ${\bf{e}}_{kl}\in\mathcal{CN}({\bf{0}},
{\bf{I}}_{N_l})$ is the i.i.d. Gaussian noise term and $\tau_{kl}$ with $0\le \tau_{kl}\le 1$ quantifies the estimation quality. Based on this model, we can reconstruct the distribution  of the channels as follows
\begin{eqnarray}
{\bf{h}}_{kl}\in\mathcal{CN}({\bf{R}}_{kl}^{1/2}(\sqrt{1-\tau_{kl}}\hat{\bf{c}}_{kl}),
{\bf{R}}_{kl}\tau_{kl}^2), \forall k, l.
\end{eqnarray}

\subsubsection{Example Three (Partial and Imperfect CSI Model)} In practice, the partial CSI knowledge
acquisition \cite{Luo_arXiv2013} (e.g., compressive CSI acquisition
\cite{Yuaning_ICC2014}) is a practical way to reduce the CSI signaling overhead
by only estimating a subset of channel links. This approach is based on the fact that the channel links between the MU and some  RAUs far away have negligible channel gains \cite{Yuaning_ICC2014,Luo_arXiv2013}, and thus the state information of these links contributes little to the performance. In the partial CSI acquisition methods, statistical channel state information is often assumed for each link. Therefore, we have mixed CSI including a subset of imperfect instantaneous CSI and statistical CSI for the other channel coefficients. Combining the above Gauss-Markov uncertainty model ({\ref{csierror}}), we can reconstruct the channel distribution for the partial and imperfect channel knowledge as follows: for the
unestimated channel links, we have $\tau_{kl}=1$, and thus the statistical knowledge is given as
${\bf{h}}_{kl}=\mathcal{CN}({\bf{0}}, {\bf{R}}_{kl})$;
for the estimated channel links with $0<\tau_{kl}<1$, the distribution of
the uncertain channel links is given by ${\bf{h}}_{kl}\in\mathcal{CN}({\bf{R}}_{kl}^{1/2}(\sqrt{1-\tau_{kl}}{\bf{c}}_{kl}),
{\bf{R}}_{kl}\tau_{kl}^2)$. In particular, $\tau_{kl}=0$ indicates that the corresponding channel coefficients are perfect.

\subsection{Stochastic Coordinated Beamforming with Probability QoS Guarantee}
The uncertainty in the available
CSI brings a new technical challenge for the system design. To guarantee
performance, we impose a probabilistic QoS constraint, specified as
follows 
\begin{eqnarray}
\label{prob}
{\rm{Pr}}\left\{{\Gamma}_{k}({\bf{v}}, {\bf{h}}_k)\ge
\gamma_{k}, \forall k\right\}\ge1-\epsilon,
\end{eqnarray}
{{where the distribution information of ${\bf{h}}_k$'s is known}}, $0<\epsilon<1$ indicates
that the system should guarantee
the QoS requirements for all the MUs {simultaneously} with probability
of at least $1-\epsilon$. The probability is calculated over all the random
vectors ${\bf{h}}_{k}$'s. The  SCB is thus formulated to minimize the total transmit
power while
satisfying the system probabilistic QoS constraint (\ref{prob}):
\begin{eqnarray}
\label{jccppp}
\mathscr{P}_{{\textrm{SCB}}}:
\mathop {\rm{minimize}}_{{\bf{v}}\in\mathcal{V}}&&\sum_{l=1}^{L}\sum_{k=1}^{K}\|{\bf{v}}_{lk}\|^2\nonumber\\
{\rm{subject~to}}&&{\rm{Pr}}\left\{{\Gamma}_{k}({\bf{v}},
{\bf{h}}_{k})\ge
\gamma_{k}, \forall k\right\}\ge1-\epsilon, 
\end{eqnarray}
which is a joint chance constrained program (JCCP)\cite{shapiro2009lectures,
hong2011sequential} and is known to be intractable in general.

\subsubsection{Problem Analysis} There are two major challenges in solving
$\mathscr{P}_{\textrm{SCB}}$. Firstly, the chance (or probabilistic) constraint
(\ref{prob}) has no closed-form expression in general
and thus is difficult to evaluate. Secondly, the convexity of the feasible set
formed by the probabilistic constraint is difficult to verify. The general
idea to handle such a constraint is to seek a \emph{safe and tractable approximation}.
``Safe" means that the feasible set formed by the approximated constraint
is a subset of  the original feasible set, while ``tractable" means that
the optimization problem over the approximated feasible set should be computationally
efficient (e.g., relaxed to a convex program). 

A natural way to form a computationally tractable approximation is the {scenario
approach} \cite{Campi_SIAM2008}. Specifically, the chance constraint (\ref{prob})
will be approximated by the following $KJ$ sampling constraints:
\begin{eqnarray}
\label{samp}
{\Gamma}_{k}({\bf{v}}, h_{k}^{j})\ge \gamma_k, 1\le j\le
J, \forall k,
\end{eqnarray} 
where $h^{j}=[h_{k}^{j}]_{1\le k\le K}, 1\le j\le J$ is a sample of
$J$ independent realizations of the random vector ${\bf{h}}$. The
SCB problem $\mathscr{P}_{\text{SCB}}$ thus can be approximated by a convex
program based on the constraints (\ref{samp}). This approach
can find a {feasible} solution with {a high probability}, for which
more details can be found in \cite{Yuaning_ICC2014}. An alternative way
is to derive an analytical upper bound for the chance constraint based on
the Bernstein-type inequality \cite{nemirovski2006convex,Vicent_TSP2013,wang2011probabilistic},
resulting in a deterministic convex optimization problem. The Bernstein approximation based
approach thus
can find a {feasible but suboptimal} solution.

Although the above methods have the advantage of computational efficiency
due to the convex approximation, the common drawback of all these algorithms
is the conservativeness due to the ``safe" approximation. Furthermore, it
is also difficult to quantify the qualities of the solutions generated by
the algorithms.   This motivates us to seek a novel approach to find a more
reliable solution to the problem $\mathscr{P}_{\textrm{SCB}}$. In this paper,
we will propose a stochastic DC programming algorithm to find the globally
optimal solution to $\mathscr{P}_{\textrm{SCB}}$ if the problem is convex
and a locally optimal solution if it is non-convex, which can be regarded
as the first attempt to guarantee the optimality for the solutions of the
JCCP (\ref{jccppp}).

\section{Stochastic DC Programming Algorithm }
\label{DCA}
In this section, we propose a stochastic
DC programming algorithm to solve the problem $\mathscr{P}_{\textrm{SCB}}$.
We will first propose a  DC programming reformulation for the problem $\mathscr{P}_{\textrm{SCB}}$,
which will then be solved by stochastic successive convex
optimization.   

\subsection{DC Programming Reformulation for the SCB Problem}
The main challenge of the SCB problem $\mathscr{P}_{\textrm{SCB}}$
is the intractable chance constraint. In order to overcome the difficulty,
we will propose a  DC programming {reformulation}
that is different from all the previous conservative approximation methods.
We first propose a DC approximation to the chance constraint (\ref{prob}).
Specifically, the QoS constraints $\Gamma_{k}({\bf{v}}, {\bf{h}}_{k})\ge\gamma_{k}$
can be rewritten as the following DC constraints \cite{horst1999dc}
\begin{eqnarray}
\label{d_DC}
d_{k}({\bf{v}},{\bf{h}}_k)\triangleq c_{k,1}({\bf{v}}_{-k},
{\bf{h}}_{k})-c_{k,2}({\bf{v}}_k,
{\bf{h}}_{k})\le 0, \forall k,
\end{eqnarray}
where ${\bf{v}}_{-k}\triangleq[{\bf{v}}_{i}]_{i\ne
k}$, and both $c_{k,1}({\bf{v}}_{-k}, {\bf{h}}_{k})\triangleq\sum_{i\ne
k}{\bf{v}}_{i}^{\sf{H}}{\bf{h}}_{k}{\bf{h}}_{k}^{\sf{H}}{\bf{v}}_{i}+\sigma_{k}^{2}$
and $c_{k,2}({\bf{v}}_k, {\bf{h}}_{k})\triangleq{1\over{\gamma_k}}{\bf{v}}_{k}^{\sf{H}}{\bf{h}}_{k}{\bf{h}}_{k}^{\sf{H}}{\bf{v}}_{k}$
are convex quadratic functions in ${\bf{v}}$. Therefore, $d_{k}({\bf{v}},
{\bf{h}}_k)$'s are DC functions in ${\bf{v}}$. Then, the chance
constraint (\ref{prob}) can be
rewritten as $f({\bf{v}})\le\epsilon$, with $f({\bf{v}})$ given by
\begin{eqnarray}
\label{indicator}
f({\bf{v}})&=&1-{\rm{Pr}}\left\{{\Gamma}_{k}({\bf{v}}, {\bf{h}}_{k})\ge
\gamma_{k}, \forall k\right\}\nonumber\\
&=&{\rm{Pr}}\left\{\left(\max_{1\le k\le K} d_{k}({\bf{v}}, {\bf{h}}_k)\right)>0\right\}\nonumber\\
&=&\mathbb{E}\left[1_{(0,+\infty)}\left(\max_{1\le k\le
K} d_{k}({\bf{v}}, {\bf{h}}_k)\right)\right],
\end{eqnarray}
where $1_{\mathcal{A}}(z)$ is an indicator of set $\mathcal{A}$. That is, $1_{\mathcal{A}}(z)=1$ if $z\in
\mathcal{A}$ and $1_{\mathcal{A}}(z)=0$, otherwise. The indicator function makes $f({\bf{v}})$
non-convex in general. 

The conventional approach to deal with the non-convex indicator function
is to approximate it
by a convex function, yielding a conservative convex approximation. For example,
using $\exp(z)\ge 1_{(0, +\infty)}(z)$ will yield the Bernstein approximation
\cite{nemirovski2006convex}. Applying $[\nu+z]^{+}/\nu\ge 1_{(0,+\infty)}(z)$,
$\nu>0$ will obtain a conditional-value-at-risk (CVaR)
type approximation \cite{nemirovski2006convex}. Although these approximations might enjoy the advantage
of being convex,
all of them are conservative and will lose optimality for the solution of
the original problem. More specifically, only the feasibility of the solutions
can be guaranteed with these approximations. 

To find a better approximation to $f({\bf{v}})$ in (\ref{indicator}), in this paper, we propose to use the following non-convex function \cite[Fig. 2]{hong2011sequential} to approximate
the indicator function $1_{(0, +\infty)}(z)$ in (\ref{indicator}):
\begin{eqnarray}
\label{DC}
\psi(z, \nu)={1\over{\nu}}[(\nu+z)^{+}-z^{+}], \nu>0,
\end{eqnarray}
which is a DC function \cite{horst1999dc} in $z$.
 Although the DC function is not convex, it does have many advantages. In
particular, Hong \emph{et al.} \cite{hong2011sequential} proposed to use
this DC function to approximate the chance constraint assuming that the functions in the chance constraint are convex, resulting in a DC program reformulation. However, we cannot
directly extend their results for our problem, since the functions $d_{k}({\bf{v}},
{\bf{h}}_{k})$'s in (\ref{d_DC}) are non-convex. Fortunately, we
can still adopt the DC function $\psi(z, \nu)$ in (\ref{DC}) to approximate
the chance constraint based on the following lemma.
\begin{lemma}[DC Approximation for the Chance Constraint]
\label{lemma_DCA} 
The non-convex function $f({\bf{v}})$ in (\ref{indicator}) has the following conservative DC approximation for any $\nu>0$,
\begin{eqnarray}
\label{fdc}
\hat{f}({\bf{v}}, \nu)&=&\mathbb{E}\left[\psi\left(\max_{1\le k\le K} d_{k}({\bf{v}},
{\bf{h}}_k), \nu\right)\right]\nonumber\\
&=&{1\over{\nu}}[u({\bf{v}}, \nu)-u({\bf{v}},
0)], \nu>0,
\end{eqnarray}
where 
\begin{eqnarray}
\label{mass}
u({\bf{v}},\nu)={\mathbb{E}}\left[\max_{1\le k\le K+1}s_{k}({\bf{v}}, \bf{h},
\nu)\right],
\end{eqnarray}
is a convex function and the convex quadratic functions $s_{k}({\bf{v}},
{\bf{h}}, \nu)$'s are given by
\begin{eqnarray}
\label{s_DC}
s_{k}({\bf{v}}, {\bf{h}}, \nu)\triangleq\nu+c_{k,1}({\bf{v}}_{-k},
{\bf{h}}_{k})+\sum_{i\ne k}c_{i,2}({\bf{v}}_{i}, {\bf{h}}_{i}), \forall k,
\end{eqnarray}
and  $s_{K+1}({\bf{v}}, {\bf{h}},
\nu)\triangleq\sum_{i=1}^{K}c_{i,2}({\bf{v}}_{i},
{\bf{h}}_{i})$ is a convex quadratic function too.
\begin{IEEEproof}
Please refer to Appendix {\ref{proof_DCA}} for details.
\end{IEEEproof}
\end{lemma}

Based on the DC approximation function $\hat{f}({\bf{v}}, \nu)$,  we propose
to solve the  following problem to approximate
the original SCB problem $\mathscr{P}_{\textrm{SCB}}$: 
\begin{eqnarray}
\label{DCR}
\mathscr{P}_{{\textrm{DC}}}:
\mathop {\rm{minimize}}_{{\bf{v}}\in\mathcal{V}}&&\sum_{l=1}^{L}\sum_{k=1}^{K}\|{\bf{v}}_{lk}\|^2\nonumber\\
{\rm{subject~to}}&& \inf_{\nu>0} \hat{f}({\bf{v}},
\nu)\le\epsilon,
\end{eqnarray}
where $\inf_{\nu>0}\hat{f}({\bf{v}}, \nu)$ is the most accurate approximation
function to $f({\bf{v}})$. Program $\mathscr{P}_{\textrm{DC}}$ is a DC program
with the convex set $\mathcal{V}$, the convex objective function, and the
DC constraint function \cite{horst1999dc}. One major advantage of the DC
approximation $\mathscr{P}_{\textrm{DC}}$ is the equivalence to the original
problem $\mathscr{P}_{\textrm{SCB}}$. That is, the DC approximation will
not lose any optimality of the solution of the SCB problem $\mathscr{P}_{\textrm{SCB}}$,
as stated in the following theorem.
\begin{theorem}[DC Programming Reformulation]
\label{theorem_DCP}
The DC programming problem $\mathscr{P}_{\textrm{DC}}$ in (\ref{DCR}) is
equivalent to the
original SCB problem $\mathscr{P}_{\textrm{SCB}}$.
\begin{IEEEproof}
Please refer to Appendix {\ref{proof_DCP}} for details.
\end{IEEEproof}
\end{theorem}

Based on this theorem, in the sequel, we focus on how to solve the problem $\mathscr{P}_{\textrm{DC}}$.

\subsection{Optimality of Joint Optimization over ${\bf{v}}$ and $\kappa$}

As the constraint in $\mathscr{P}_{\textrm{DC}}$ itself is an optimization
problem, it is difficult to be solved directly. To circumvent this difficulty, by observing that $\hat{f}({\bf{v}}, {\nu})$ is nondecreasing in $\nu$
for $\nu>0$, as indicated in (\ref{nonde}), one way is  to solve the following
$\kappa$-approximation problem \cite{hong2011sequential}
\begin{eqnarray}
\label{fixeddc}
\mathop {\rm{minimize}}_{{\bf{v}}\in\mathcal{V}}&&\sum_{l=1}^{L}\sum_{k=1}^{K}\|{\bf{v}}_{lk}\|^2\nonumber\\
{\rm{subject~to}}&& u({\bf{v}},\kappa)-u({\bf{v}},0)\le\kappa\epsilon,
\end{eqnarray}
for any fixed small enough parameter $\kappa>0$ to approximate the original problem $\mathscr{P}_{\textrm{DC}}$. However, an extremely small $\kappa$ might cause numerical stability
issues and might require more time to solve the subproblems that will be developed later \cite{hong2011sequential}.

We notice that, by regarding $\kappa$ as an optimization variable, problem (\ref{fixeddc}) is still a DC program, as the function $\mu({\bf{v}}, \kappa)$ is jointly convex in $({\bf{v}}, \kappa)$. Therefore, we propose to solve the
following joint approximation 
optimization problem by treating $\kappa$ as an optimization variable
\begin{eqnarray}
\tilde{\mathscr{P}}_{\textrm{DC}}:
\mathop {\rm{minimize}}_{{\bf{v}}\in\mathcal{V}, \kappa>0}&&\sum_{l=1}^{L}\sum_{k=1}^{K}\|{\bf{v}}_{lk}\|^2\nonumber\\
{\rm{subject~to}}&& [u({\bf{v}},\kappa)-\kappa\epsilon]-u({\bf{v}},0)\le 0.
\end{eqnarray}
The following proposition implies that the joint approximation problem $\tilde{\mathscr{P}}_{\textrm{DC}}$ can enhance the performance of problem (\ref{fixeddc}).
\begin{proposition}[Effectiveness of Joint Approximation]
\label{projoint}
Denote the optimal value of the problem (\ref{fixeddc}) with a fixed $\kappa=\hat{\kappa}$ and that of the problem $\tilde{\mathscr{P}}_{\textrm{DC}}$ as $V^{\star}(\hat{\kappa})$ and $\tilde{V}^{\star}$, respectively, then we have $ \tilde{V}^{\star}\le V^{\star}(\hat{\kappa})$.
\begin{IEEEproof}
Define the feasible region of problem $\tilde{\mathscr{P}}_{{\textrm{DC}}}$
as 
\begin{eqnarray}
\mathcal{D}\triangleq\{{\bf{v}}\in\mathcal{V}, \kappa>
0: [u({\bf{v}},\kappa)-\kappa\epsilon]-u({\bf{v}},0)\le 0\}.
\end{eqnarray}
 The projection
of ${\mathcal{D}}$
on the set $\mathcal{V}$ is given by 
\begin{eqnarray}
\bar{\mathcal{D}}=\{{\bf{v}}\in{\mathcal{D}}:
\exists \kappa> 0, {\textrm{s.t.}} ({\bf{v}}, \kappa)\in\mathcal{D}\}.
\end{eqnarray}
Therefore, fixing $\kappa=\hat{\kappa}$, any feasible solution in problem (\ref{fixeddc}) belongs to the set $\mathcal{D}$. Therefore, the feasible set of the optimization problem  (\ref{fixeddc}) is a subset of $\bar{\mathcal{D}}$. As a result, solving $\tilde{\mathscr{P}}_{\textrm{DC}}$ can achieve a smaller minimum value with a larger feasible region.
\end{IEEEproof} 
\end{proposition}

Define the {deviation}
of a given set $\mathcal{A}_1$ from another set $\mathcal{A}_2$ as \cite{shapiro2009lectures}
\begin{eqnarray}
\mathbb{D}(\mathcal{A}_1, \mathcal{A}_2)=\sup_{x_1\in \mathcal{A}_1}\left(\inf_{x_2\in
\mathcal{A}_2}\|x_1-x_2\|\right),
\end{eqnarray}
then we have the following theorem indicating the optimality of the joint approximation program $\tilde{\mathscr{P}}_{\textrm{DC}}$.
\begin{theorem}[Optimality of Joint Approximation]
\label{dcapp}
Denote the set of the optimal solutions and optimal values of problems $\tilde{\mathscr{P}}_{\textrm{DC}}$, $\mathscr{P}_{\textrm{SCB}}$ and the problem (\ref{fixeddc}) with a fixed $\kappa=\hat{\kappa}$ as ($\tilde{\mathcal{P}}^{\star}, \tilde{V}^{\star}$),
\!($\mathcal{P}^{\star}, V^{\star}$) and ($\mathcal{P}^{\star}(\hat{\kappa}), V^{\star}(\hat{\kappa})$), respectively, we have 
\begin{eqnarray}
\lim_{\hat{\kappa}\searrow
0}(V^{\star}(\hat{\kappa})-\tilde{V}^{\star})=\lim_{\hat{\kappa}\searrow
0}(V^{\star}(\hat{\kappa})-V^{\star})=0,
\end{eqnarray}
and 
\begin{eqnarray}
\lim_{\hat{\kappa}\searrow
0} \mathbb{D}(\mathcal{P}^{\star}(\hat{\kappa}), \tilde{\mathcal{P}}^{\star})=\lim_{\hat{\kappa}\searrow
0} \mathbb{D}(\mathcal{P}^{\star}(\hat{\kappa}), {\mathcal{P}}^{\star})=0.
\end{eqnarray}

\begin{IEEEproof}
Based on Proposition {\ref{projoint}}, the proof follows \cite[Theorem 2]{hong2011sequential}. \end{IEEEproof}
\end{theorem}

Based on Theorem {\ref{dcapp}}, we can thus focus on solving the program $\tilde{\mathscr{P}}_{\textrm{DC}}$. Although $\tilde{\mathscr{P}}_{\textrm{DC}}$
is still a non-convex DC program, it has the algorithmic advantage, as will
be presented in the next subsection.

\subsection{Successive Convex Approximation Algorithm}
In this subsection, we will present a successive convex approximation algorithm
\cite{hong2011sequential,Luo2013unified} to solve the non-convex joint approximation
program $\tilde{\mathscr{P}}_{\textrm{DC}}$. We will prove in Theorem {\ref{theorem_CON}} that this algorithm still preserves the optimality properties, i.e., achieving the Karush-Kuhn-Tucker (KKT) pair of the non-convex program $\tilde{\mathscr{P}}_{\textrm{DC}}$. The main idea  is to upper bound
the non-convex DC constraint function in $\tilde{\mathscr{P}}_{\textrm{DC}}$
by a convex function at each iteration. Specifically, at the $j$-th iteration,
given
the vector $({\bf{v}}^{[j]}, \kappa^{[j]})\in\mathcal{D}$, for the convex function $u({\bf{v}}, 0)$, we have 
\begin{eqnarray}
u({\bf{v}}, 0)\ge u({\bf{v}}^{[j]}, 0)+2\langle{ \nabla_{{\bf{v}}^{*}}u({\bf{v}}^{[j]},
0), {\bf{v}}-{\bf{v}}^{[j]}}\rangle, 
\end{eqnarray}
where $\langle {\bf{a}}, {\bf{b}}\rangle\triangleq\mathfrak{R} ({\bf{a}}^{\sf{H}}{\bf{b}})$
for any ${\bf{a}}, {\bf{b}}\in\mathbb{C}$ and the gradient of function $u({\bf{v}},0)$
is
given as follows.

\begin{lemma}
\label{lemma_GRAD}
The complex gradient of $u({\bf{v}}, 0)$ with
respect to ${\bf{v}}^{*}$ (the complex conjugate of $\bf{v}$)
is given by 
\begin{eqnarray}
\nabla_{{\bf{v}}^{*}}u({\bf{v}},0)= \mathbb{E}[\nabla_{{\bf{v}}^{*}}s_{
k^{\star}}({\bf{v}},{\bf{h}},0)],
\end{eqnarray}
where $k^{\star}=\arg\max\limits_{1\le k\le K+1}s_{k}({\bf{v}}, {\bf{h}},0)$,
and $\nabla_{{\bf{v}}^{*}}s_{k}({\bf{v}},{\bf{h}},0)=[{\boldsymbol{\nu}}_{k,i}]_{1\le
i\le K}(1\le k\le K)$ with ${\boldsymbol{\nu}}_{k,i}\in\mathbb{C}^{N}$ given
by
\begin{eqnarray}
{\boldsymbol{\nu}}_{k,i} = \left\{ \begin{array}{ll}
\left( {\bf{h}}_{k}{\bf{h}}_{k}^{\sf{H}}+{1\over{\gamma_{i}}} {\bf{h}}_{i}{\bf{h}}_{i}^{\sf{H}}\right){\bf{v}}_{i},
& \textrm{if $i\ne k$}, 1\le k\le K,\\
 0, & \textrm{otherwise},\nonumber
  \end{array} \right.
\end{eqnarray}
and $\nabla_{{\bf{v}}^{*}}s_{K+1}({\bf{v}},{\bf{h}},\kappa)=[{\boldsymbol{\nu}}_{K+1,i}]_{1\le
i\le K}$ with ${\boldsymbol{\nu}}_{K+1,i}={1\over{\gamma_{i}}} {\bf{h}}_{i}{\bf{h}}_{i}^{\sf{H}}{\bf{v}}_{i},
\forall i$. Furthermore, the gradient of $u({\bf{v}}, 0)$ with
respect to $\kappa$ is zero, as $\kappa=0$ is a constant in the function $u({\bf{v}}, 0)$.
\begin{IEEEproof}
Please refer to Appendix {\ref{proof_GRAD}} for details. 
\end{IEEEproof}
\end{lemma}
Therefore, at the $j$-th iteration, the non-convex DC constraint function $[u({\bf{v}},\kappa)-\kappa\epsilon]-u({\bf{v}},0)$ in $\tilde{\mathscr{P}}_{\textrm{DC}}$
can be upper bounded by the 
convex function $l({\bf{v}}, \kappa;{\bf{v}}^{[j]}, \kappa^{[j]})-\kappa\epsilon$ with  
\begin{eqnarray}
\label{linear}
&&l({\bf{v}}, \kappa;{\bf{v}}^{[j]}, \kappa^{[j]})\nonumber\\
&&=u({\bf{v}},
\kappa)- u({\bf{v}}^{[j]},
0)- 2\langle{ \nabla_{{\bf{v}}^{*}}u({\bf{v}}^{[j]},
0), {\bf{v}}-{\bf{v}}^{[j]}}\rangle.
\end{eqnarray}
 Based on the convex approximation (\ref{linear}) to the DC constraint in
$\tilde{\mathscr{P}}_{\textrm{DC}}$, we
will then solve the following stochastic convex programming problem at
the $j$-th iteration:
\begin{eqnarray}
\label{sdc1}
\!\!\!\!\!\!\tilde{\mathscr{P}}_{{\textrm{DC}}}({\bf{v}}^{[j]}, \kappa^{[j]}):
\mathop {\rm{minimize}}_{{\bf{v}}\in\mathcal{V}, \kappa>0}&&\sum_{l=1}^{L}\sum_{k=1}^{K}\|{\bf{v}}_{lk}\|^2\nonumber\\
{\rm{subject~to}}&&l({\bf{v}}, {\kappa}; {\bf{v}}^{[j]}, \kappa^{[j]})-\kappa\epsilon\le0.
\end{eqnarray}

The proposed stochastic DC programming algorithm to the SCB problem $\mathscr{P}_{\textrm{SCB}}$
is thus presented
in Algorithm 1.

\begin{algorithm}
\caption{Stochastic DC Programming Algorithm}
\textbf{Step 0:} Find the initial solution $({\bf{v}}^{[0]}, \kappa^{[0]})\in \mathcal{D}$
and set the iteration counter $j=0$;\\
\textbf{Step 1:} {\textbf{If}} (${\bf{v}}^{[j]}, \kappa^{[j]})$ satisfies the termination
criterion,
{\textbf{go to End}};\\
\textbf{Step 2:} Solve problem $\tilde{\mathscr{P}}_{\textrm{DC}}({\bf{v}}^{[j]},
{\kappa^{[j]}})$ and obtain the optimal solution $({\bf{v}}^{[j+1]}, \kappa^{[j+1]})$;\\
\textbf{Step 3:} Set $j=j+1$ and {\textbf{go to Step 1}}; \\
\textbf{End}.
\end{algorithm}

Based on Theorem {\ref{dcapp}} on the optimality of the joint approximation,
the convergence of the stochastic DC programming algorithm is presented in
the following theorem, which reveals the main advantage compared with all
the previous algorithms for the JCCP problem, i.e., it guarantees optimality.
\begin{theorem}[Convergence of Stochastic DC Programming]
\label{theorem_CON}
Denote $\{{\bf{v}}^{[j]}, \kappa^{[j]}\}$ as the sequence generated by the stochastic DC
programming algorithm. Suppose that the limit of the sequence exists, i.e.,
$\lim_{j\rightarrow +\infty}({\bf{v}}^{[j]}, \kappa^{[j]})=({\bf{v}}^{{\star}}, \kappa^{\star})$, which satisfies
the Slater's condition\footnote{Slater's condition is a commonly used constraint
qualification to ensure the existence of KKT pairs in convex optimization\cite{boyd2004convex}. },  then ${\bf{v}}^{\star}$ is the
globally optimal solution of the SCB problem
$\mathscr{P}_{\textrm{SCB}}$ if it is convex; otherwise, ${\bf{v}}^{\star}$
is a locally optimal solution. Furthermore, $\kappa$ converges to zero for most scenarios, except that 
\begin{eqnarray}
\label{zzc1}
{\rm{Pr}}\left\{\max_{1\le k\le K}d_{k}({\bf{v}}^{\star}, {\bf{h}})\in(-\kappa^{\star},
0]\right\}=0,
\end{eqnarray}
if $\kappa^{\star}\ne 0$.
\begin{IEEEproof}
Please refer to Appendix \ref{proof_CON} for details.
\end{IEEEproof}
\end{theorem}

Based on Theorem {\ref{theorem_CON}}, in the sequel, we focus on how to efficiently implement the stochastic DC programming algorithm.

\subsection{Sample Average Approximation Method for the Stochastic DC Programming Algorithm}

In order to implement the stochastic DC programming algorithm, we need to address the problem on  how to solve the stochastic convex program $\tilde{\mathscr{P}}_{\textrm{DC}}({\bf{v}}^{[j]}, \kappa^{[j]})$ (\ref{sdc1}) efficiently at each iteration.

We propose to use the sample average approximation (SAA) based
algorithm \cite{shapiro2009lectures} to solve the stochastic
convex problem $\tilde{\mathscr{P}}_{{\textrm{DC}}}({\bf{v}}^{[j]}, \kappa^{[j]})$ at the $j$-th
iteration. Specifically, the SAA estimate of $u({\bf{v}}, \kappa)$
is
given by
\begin{eqnarray}
\bar{u}({\bf{v}}, \kappa)={1\over{M}}\sum_{m=1}^{M}\max_{1\le k\le K+1}s_{k}({\bf{v}},
h^{m}, \kappa),
\end{eqnarray}
where $h^{m} (1\le m\le M)$ is a sample of $M$ independent realizations
of the random vector ${\bf{h}}$.
Similarly, the SAA estimate of the gradient $\nabla_{{\bf{v}}^{*}}u({\bf{v}},0)$
is given by
\begin{eqnarray}
\label{GRAD_estimator}
\bar\nabla_{{\bf{v}}^{*}}u({\bf{v}},0)={1\over{M}}\sum_{m=1}^M\nabla_{{\bf{v}}^{*}}s_{
k_{m}^{\star}}({\bf{v}},{{h}}^m, 0),
\end{eqnarray}
where $k_{m}^{\star}=\arg\max\limits_{1\le k\le K+1} s_{k}({\bf{v}}, h^{m},
0)$. Therefore,
the SAA estimate of the convex function $l({\bf{v}}, \kappa;{\bf{v}}^{[j]}, \kappa^{[j]})$
(\ref{linear}) is given by
\begin{eqnarray}
\label{linear_est}
&&\bar l({\bf{v}}, \kappa;{\bf{v}}^{[j]}, \kappa^{[j]})\nonumber\\
&&=\bar u({\bf{v}},
\kappa)- \bar u({\bf{v}}^{[j]},
0)- 2\langle{ \bar\nabla_{{\bf{v}}^{*}}u({\bf{v}}^{[j]},
0), {\bf{v}}-{\bf{v}}^{[j]}}\rangle,
\end{eqnarray}
which is jointly convex in $\bf{v}$ and $\kappa$.
We will thus solve the following SAA based convex optimization
problem
\begin{eqnarray}
\!\!\!\!\!\!\!\!\!\bar{\mathscr{P}}_{{\textrm{DC}}}({\bf{v}}^{[j]}, \kappa^{[j]}; M):
\mathop {\rm{minimize}}_{{\bf{v}}\in\mathcal{V}, \kappa>0}\!&&\sum_{l=1}^{L}\sum_{k=1}^{K}\|{\bf{v}}_{lk}\|^2\nonumber\\
{\rm{subject~to}}&&\!\bar l({\bf{v}}, \kappa;{\bf{v}}^{[j]}, \kappa^{[j]})-\kappa\epsilon\le0,
\end{eqnarray}
to approximate the stochastic convex optimization problem $\tilde{\mathscr{P}}_{{\textrm{DC}}}({\bf{v}}^{[j]}, \kappa^{[j]})$,
which can be reformulated as the following convex quadratically constraint
quadratic program (QCQP) \cite{boyd2004convex}:
\begin{eqnarray}
\label{qcqp}
\!\!\!\!\!\!\!\mathscr{P}_{\textrm{QCQP}}^{[j]}:
\mathop {\rm{minimize}}_{{\bf{v}}\in\mathcal{V}, \kappa> 0, {\bf{x}}}&&\sum_{l=1}^{L}\sum_{k=1}^{K}\|{\bf{v}}_{lk}\|^2\nonumber\\
{\rm{subject~to}}&&{1\over{M}}\sum_{m=1}^{M}x_{m}-\bar u({\bf{v}}^{[j]},
0)\nonumber\\
&&-2\langle{\bar
\nabla_{{\bf{v}}^{*}}u({\bf{v}}^{[j]},
0), {\bf{v}}-{\bf{v}}^{[j]}}\rangle\le\kappa\epsilon \nonumber\\
&&s_{k}({\bf{v}},
h^{m}, \kappa)\le x_m, x_{m}\ge 0, \forall k, m,
\end{eqnarray}
 which can then be solved efficiently using the interior-point method \cite{boyd2004convex},
where ${\bf{x}}=[x_{m}]_{1\le m\le M}\in\mathbb{R}^M$ is the collection of
the slack variables. 

The following theorem indicates that the SAA based program $\bar{\mathscr{P}}_{{\textrm{DC}}}({\bf{v}}^{[j]}, \kappa^{[j]}; M)$ for the stochastic convex optimization $\tilde{\mathscr{P}}_{{\textrm{DC}}}({\bf{v}}^{[j]}, \kappa^{[j]})$
will not lose any optimality in the asymptotic regime.

\begin{theorem}
\label{theorem_est}
Denote the set of the optimal solutions and optimal values of problems $\mathscr{P}_{{\textrm{DC}}}({\bf{v}}^{[j]}, \kappa^{[j]})$
and $\bar{\mathscr{P}}_{{\textrm{DC}}}({\bf{v}}^{[j]}, \kappa^{[j]}; M)$
as $(\mathcal{\mathcal{P}}^{\star}({\bf{v}}^{[j]}, \kappa^{[j]}), V^{\star}({\bf{v}}^{[j]}, \kappa^{[j]}))$ and $(\mathcal{P}_M^{\star}({\bf{v}}^{[j]}, \kappa^{[j]}), V_M^{\star}({\bf{v}}^{[j]}, \kappa^{[j]}))$,
respectively, then we have
\begin{eqnarray} 
\mathbb{D}(\mathcal{P}_M^{\star}({\bf{v}}^{[j]}, \kappa^{[j]}),
\mathcal{P}^{\star}({\bf{v}}^{[j]}, \kappa^{[j]}))\rightarrow 0,
\end{eqnarray}
and 
\begin{eqnarray}
V_{M}^{\star}({\bf{v}}^{[j]}, \kappa^{[j]})\rightarrow V^{\star}({\bf{v}}^{[j]}, \kappa^{[j]}),
\end{eqnarray}
with
probability one, as the sample size increases, i.e., as $M\rightarrow +\infty$.
\begin{IEEEproof}
Please refer to Appendix \ref{proof_est} for details.
\end{IEEEproof}
\end{theorem}

Based on Theorems 1-4, we conclude that the proposed stochastic DC programming
algorithm converges to the globally optimal solution of the SCB problem if
it is convex and to a locally optimal solution if the problem is non-convex,
in the asymptotic regime, i.e., $M\rightarrow +\infty$.  

\subsection{Complexity Analysis and Discussions}
To implement the stochastic DC programming
algorithm, at each iteration, we need to
solve the convex QCQP program $\mathscr{P}_{\textrm{QCQP}}^{[j]}$  with $m=(L+KM+1)$ ($M$ is the number of independent realizations of the random
vector ${\bf{h}}$) constraints and $n=(NK+M+1)$ optimization variables. The convex QCQP problem can be solved with a worst-case complexity of $\mathcal{O}((mn^2+n^3)m^{1/2}\log(1/\varepsilon))$ given a solution accuracy $\varepsilon>0$ using the interior-point method \cite{nesterov1994interior}. As the Monte Carlo sample size $M$  could be very large in order to reduce the approximation bias \cite{hong2011sequential}, the computational complexity of the stochastic DC programming algorithm could be higher than other deterministic approximation methods, e.g., the Bernstein approximation method.  

{{In order to further improve the computational
efficiency of the stochastic DC programming algorithm, other approaches can
be explored (e.g., the {alternating direction method of multipliers}
(ADMM) method \cite{boyd2011distributed}) to solve the large-scale conic
program $\mathscr{P}_{\textrm{QCQP}}^{[j]}$ in (\ref{qcqp})
at each iteration. This is an on-going research topic, and we will leave
it as our future work. }} 

Furthermore, as the stochastic DC programming algorithm only requires distribution information of the
random vector ${\bf{h}}$ to generate the Monte Carlo samples, this approach can be widely applied for any channel uncertainty model. {{As the proposed stochastic
DC programming algorithm provides optimality guarantee, it can serve as the performance benchmark in various beamforming design problems with CSI uncertainty and probabilistic
QoS guarantees, and thus it will find wide applications
in future wireless networks.}}

\section{Simulation Results}
\label{sim}
In this section, we simulate the proposed stochastic DC algorithm for coordinated beamforming design. We consider
the following channel model for the link between the $k$-th user and the
$l$-th RAU  \cite{WeiYu_WC10,Jeffrey_2011mimo}:
\begin{eqnarray}
{\bf{h}}_{kl}&=&\underbrace{10^{-L(d_{kl})/20}\sqrt{\varphi_{kl}s_{kl}}}_{D_{kl}}\left(\sqrt{1-\tau_{kl}^2}{\hat{\bf{c}}}_{kl}+\tau_{kl}{\bf{e}}_{kl}\right)\nonumber\\
&=&\sqrt{1-\tau_{kl}^2}D_{kl}\hat{\bf{c}}_{kl}+\tau_{kl}D_{kl}{\bf{e}}_{kl}, \forall k, l,
\end{eqnarray}
where $L(d_{kl})$ is the path-loss at distance $d_{kl}$, as given in \cite[Table I]{Yuanming_TWC2014}, $s_{kl}$ is the shadowing coefficient, $\varphi_{kl}$ is
the antenna gain, $\hat{\bf{c}}_{kl}\in\mathcal{CN}({\bf{0}}, {\bf{I}}_{N_l})$ is the estimated imperfect small-scale fading
coefficient and ${\bf{e}}_{kl}$ is the CSI error. We assume  that the BBU
pool can accurately track the large-scale fading coefficients $D_{kl}$'s
\cite{Luo_arXiv2013}. The error vector is modeled as
${\bf{e}}_{kl}\in\mathcal{CN}({\bf{0}}, {\bf{I}}_{N_l})$. The parameters
$\tau_{kl}$'s depend on the CSI acquisition schemes, e.g., channel
estimation errors using MMSE.
We use the standard cellular network parameters as shown in \cite[Table I]{Yuanming_TWC2014}.
 The maximum outage probability that the system can tolerate is set as $\epsilon=0.1$.  The proposed stochastic DC programming algorithm
will stop if the difference between the objective values of $\tilde{\mathscr{P}}_{{\textrm{DC}}}({\bf{v}}^{[j]}, \kappa^{[j]})$ (\ref{sdc1}) of two consecutive
iterations is less than $10^{-4}$. 

{{The proposed stochastic DC programming algorithm is compared to the following two
algorithms:
\begin{itemize}
\item {\textbf{The scenario approach}}: The main idea of this algorithm is
to approximate the probabilistic QoS constraint by multiple ``sampling" QoS
constraints \cite{Campi_SIAM2008,Angela_JSAC2013}. This algorithm can only find a feasible solution for problem $\mathscr{P}_{\textrm{SCB}}$ with a high probability. Please refer to \cite{Yuaning_ICC2014} for more details. 

\item {\textbf{The Bernstein approximation method}}: The main idea of this algorithm is to use the Bernstein-type inequality to find a closed-form approximation for the chance constraint (\ref{prob}) \cite{Vicent_TSP2013,wang2011probabilistic}. The original stochastic optimization problem $\mathscr{P}_{\textrm{SCB}}$ can be conservatively approximated by a deterministic optimization problem. Therefore, the computational complexity of the deterministic approximation method is normally much lower than the Monte Carlo approaches, e.g., the scenario approach and the stochastic DC programming algorithm. Nevertheless, the Bernstein approximation method
can also only find a feasible but suboptimal solution, and the conservativeness
of this method is difficult to quantify. Moreover,  to derive closed-form expressions, the Bernstein approximation
method restricts the distribution of 
the random vector ${\bf{h}}$ to be complex Gaussian distribution. Therefore, this method is not robust against
the distribution of the random vector $\bf{h}$. \end{itemize}
}}

Due to the computational complexity of solving large-size sample problems for both the stochastic DC programming algorithm and the scenario approach, we only consider a simple and particular network realization to demonstrate the performance benchmarking capability of the proposed stochastic DC programming algorithm. Specifically, consider a network with $L=5$ single-antenna
RAUs and $K=3$ single-antenna MUs uniformly and independently distributed
in the square region $[-400, 400]\times [-400, 400]$ meters. In this scenario,
we consider a mixed CSI uncertainty model \cite{Yuaning_ICC2014,Luo_arXiv2013},
i.e., partial and imperfect CSI.
Specifically, for MU $k$, we set $\tau_{kn}=0.01, \forall n\in\Omega_{k}$ (i.e.,
the obtained channel coefficients are {imperfect}) and $\tau_{kn}=1,
\forall\neq\Omega_{k}$, where $\Omega_{k}$ includes the indices of the 2 largest
entries of the vector consisting of all the large-scale fading coefficients
for MU $k$. That is, only $40\%$ of the channel coefficients are  obtained in this
scenario.
The QoS requirements are set as $\gamma_{k}=3
{\rm{dB}}, \forall k$. The sample size for the scenario approach
is 308 \cite{Campi_SIAM2008} and for the stochastic DC programming algorithm
it is $1000$. The simulated channel data is given in (\ref{data}), where $\hat{\bf{H}}=[D_{kl}\hat{\bf{c}}_{kl}]$ and ${\bf{D}}=[D_{kl}]$. In the following, we will illustrate the convergence,
conservativeness, stability and performance gains of the stochastic DC programming algorithm. 
\begin{figure*}
\begin{eqnarray}
\label{data}
{{\bf{\hat{{H}}}} =
\left[ \begin{array}{lll}
  -2.2377 + 0.9643i ~~& -1.0311 + 2.0312i  ~~&3.6613 +11.3275i  \\
 -0.5723 - 0.1608i  ~~& 8.4672 +19.4963i ~~&-0.0046 - 1.3821i  \\
 28.8976 -13.2169i  ~~&4.3453 -10.1453i ~~& 1.6451 - 4.8108i \\
-1.6776 + 1.2600i   ~~&  -2.6659 - 2.0050i  ~~~& 42.9821 - 5.6807i \\
3.4623 - 2.0804i  ~~& 4.1266 + 1.8647i ~~& -2.3121 + 1.3415i 
\end{array} \right]}, {\bf{D} =
\left[ \begin{array}{lll}
2.7963 ~~& 4.4546  ~~& 26.8928  \\
 2.4794 ~~& 9.5564 ~~& 1.9145  \\
 29.9654 ~~& 24.3376 ~~& 13.8270 \\
2.1076   ~~& 4.0912 ~~& 38.7970 \\
2.8683 ~~& 3.9187 ~~& 3.5856 
\end{array} \right]}.
\end{eqnarray}
\end{figure*}

\subsection{Stability of the Algorithms}
As both the stochastic DC programming and scenario approach  use Monte Carlo samples to obtain the solutions, the corresponding solutions should depend on the particular samples. Therefore, it is essential to investigate the stability of solutions obtained by the stochastic algorithms. We thus run the algorithms 50 replications with different Monte Carlo samples for each replication to illustrate the stability of the algorithms. 
\begin{figure}[t]
  \centering
  \includegraphics[width=0.95\columnwidth]{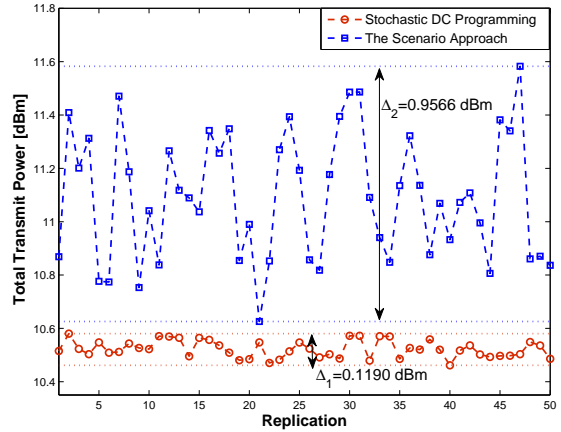}
 \caption{Optimal value versus different Monte Carlo  replications.}
 \label{stable11}
 \end{figure}
 
 \begin{figure}[t]
  \centering
  \includegraphics[width=0.95\columnwidth]{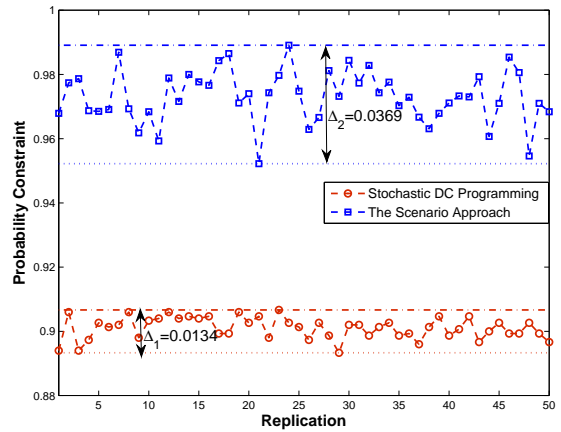}
 \caption{Probability constraint versus different Monte Carlo  replications.}
  \label{stableprob11}
 \end{figure}

From Fig. {\ref{stable11}} and Fig. {\ref{stableprob11}}, we can see that the solutions and the estimated probability constraints obtained from the stochastic DC programming algorithm are very stable, as they converge to a similar solution. In particular, the average total transmit power is 10.5228 dBm, with the lowest being 10.4614 dBm and the highest being 10.5804 dBm. The corresponding average probability constraint is 0.9010, with the range of 0.8933 to  0.9067.

However, the solutions and  the estimated probability constraints obtained from the scenario approach drastically differ from replication to replication due to the randomness in the Monte Carlo samples. In particular, the average total transmit power is 11.1004
dBm, with the lowest being 10.6260 dBm and the highest being 11.5826 dBm.
The corresponding average probability constraint is 0.9731, and is in the
range between 0.9522 and  0.9891.

We can see that  the stochastic DC programming algorithm can achieve a lower
transmit power than the scenario approach on average. The scenario approach yields a much more conservative approximation for the probability constraint. Furthermore, the performance of the scenario approach cannot be improved by increasing the sampling size  as this will cause more conservative solutions. This is in contrast to the proposed stochastic DC programming algorithm, as Theorem {\ref{theorem_est}}  indicates that more samples can improve the Monte Carlo approximation performance and most Monte Carlo approach based stochastic  algorithms possess such a property.

Finally, the average value of the parameter $\kappa$ is $1.5\times 10^{-3}$ and is in the rang between $7.8\times 10^{-4}$ and $2.6\times 10^{-3}$
 when the stochastic
DC programming algorithm terminates. This justifies the conclusion that the parameter $\kappa$ will converge to zero as  presented
in Theorem {\ref{theorem_CON}}.

\subsection{Convergence of the Stochastic DC Programming Algorithm}
\begin{figure}[t]
  \centering
  \includegraphics[width=0.95\columnwidth]{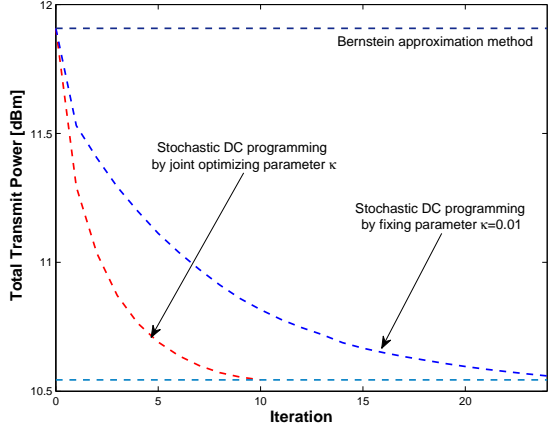}
 \caption{Convergence of the stochastic DC programming algorithm.}
 \label{sim_conver}
 \end{figure}

 We report a typical performance on the convergence of the stochastic DC programming algorithm, as shown in Fig. {\ref{sim_conver}},
with the initial point being the solution from the Bernstein approximation method.
This figure shows that the convergence rate of the proposed stochastic DC
programming is very fast for the simulated scenario. We can see that  the stochastic DC programming algorithm can achieve a much lower
transmit power than the Bernstein approximation
method. This figure also demonstrates the effectiveness of jointly optimizing over the parameter $\kappa$ and beamforming vector ${\bf{v}}$, as this can significantly improve the convergence rate. Furthermore, the parameter $\kappa$ is $1.3\times10^{-3}$ when the proposed stochastic DC programming algorithm terminates under this scenario.  

\subsection{Conservativeness of the Algorithms}
\begin{figure}[t]
  \centering
  \includegraphics[width=0.95\columnwidth]{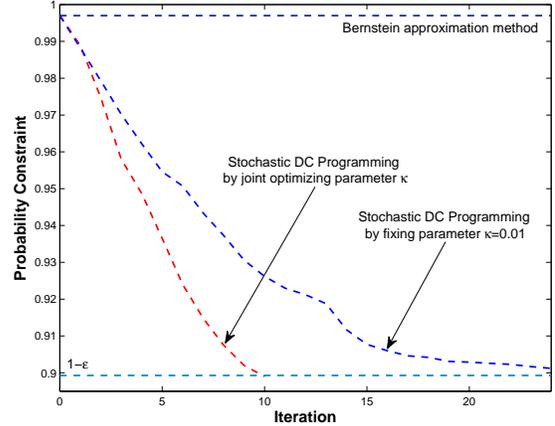}
 \caption{Probability Constraint.}
 \label{probability}
 \end{figure}

We also report the typical performances of all the algorithms on the  conservativeness of approximating probability constraints in the SCB problem under the same scenario as the above subsection. The estimated probability constraint in $\mathscr{P}_{\textrm{SCB}}$ is shown in Fig. {\ref{probability}}, which is 0.988 using the Bernstein approximation. On the other hand, for the stochastic DC programming algorithm, we can see that the probability constraint becomes tight
when it terminates, and thus the Bernstein approximation is too conservative. This coincide with the fact that the suboptimal algorithms only seek  conservative approximations to the chance constraint.     

\section{Conclusions and Future Works} 
\label{con_dis}
This paper presented a generic stochastic coordinated beamforming framework for the optimal transmission strategy design with a probabilistic model for the CSI uncertainty. This framework frees us from the structural modeling assumptions and distribution types assumptions for the uncertain channel knowledge, thus it provides modeling flexibility. With the optimality guarantee, the proposed stochastic DC programming algorithm can serve as the benchmark for evaluating suboptimal and heuristic algorithms. The benchmarking capability was demonstrated numerically in terms of conservativeness, stability and optimal  values by comparing with the Bernstein approximation method and scenario approach. Furthermore, the proposed algorithm has a better convergence rate by jointly optimizing the approximation parameter $\kappa$. As the proposed stochastic
DC programming algorithm provides optimality guarantee, we believe this algorithm
can be applied in various beamforming design problems with probabilistic
QoS guarantees due to the CSI uncertainty, and it will find wide applications
in future wireless networks.

Several future research directions are listed as follows:
\begin{itemize}
\item Although our framework only requires the distribution information of the uncertain channel knowledge, so as to generate Monte Carlo samples for the stochastic DC programming algorithm, it might be challenging to obtain the exact information in some scenarios. Therefore, one may either seek more sophisticated measuring methods to estimate the distribution information or  adopt the distributionally robust optimization approaches to deal with the ambiguous distributions, e.g., \cite{Ye_2010distributionally}.

\item The main drawback of the stochastic DC programming algorithm is the highly computational complexity with the sample problem $\mathscr{P}_{\textrm{QCQP}}^{[j]}$ at each iteration, one may either resort to ADMM \cite{boyd2011distributed} based algorithms to solve the large-sized sample problem in parallel or reduce the optimization dimensions by fixing the directions of the beamformers and only optimizing the transmit power allocation (e.g., in\cite{Xiaodong_2014power}, the corresponding power allocation problem is a linear program and can be solved with a much lower computational complexity).
\end{itemize}

\appendices

\section{Proof of Lemma \ref{lemma_DCA}}
\label{proof_DCA}
For simplicity, we denote $c_{k,1}({\bf{v}})\triangleq c_{k,1}({\bf{v}}_{-k},
{\bf{h}}_{k})$, $c_{k,2}({\bf{v}})\triangleq c_{k,2}({\bf{v}}_k,
{\bf{h}}_k)$ and $d_{k}({\bf{v}})\triangleq d_{k}({\bf{v}}, {\bf{h}}_k)$.
For any ${\bf{v}}\in\mathcal{V}$, $\forall k$, $d_{k}({\bf{v}})=c_{k,1}({\bf{v}})-c_{k,2}({\bf{v}})$ is a DC function on $\mathcal{V}$, as both $c_{k,1}({\bf{v}})$ and $c_{k,2}({\bf{v}})$ are convex functions of ${\bf{v}}$. For any  $\nu>0$,
we first prove that the following function   
\begin{eqnarray}
&&\psi\left(\max_{1\le k\le K}d_{k}({\bf{v}}), \nu\right)=\nonumber\\
&&{1\over{\nu}}\left[\left(\nu+\max_{1\le
k\le K} d_{k}({\bf{v}})\right)^{+}-\left(\max_{1\le k\le K}d_{k}({\bf{v}})\right)^{+}\right],
\end{eqnarray} 
is also a DC function. The function $d_{k}({\bf{v}})$ can be rewritten as
\begin{eqnarray}
\label{dd}
d_{k}({\bf{v}})=c_{k,1}({\bf{v}})+\sum_{i\ne k}c_{i,2}({\bf{v}})-\sum_{i=1}^{K}c_{i,2}({\bf{v}}).
\end{eqnarray}
Therefore, the following function
\begin{eqnarray}
\label{bb3}
&&\max_{1\le k\le K} d_{k}({\bf{v}})=\nonumber\\
&&\underbrace{\max_{1\le k\le K}\left\{c_{k,1}({\bf{v}})+\sum_{i\ne
k}c_{i,2}({\bf{v}})\right\}}_{C_{1}({\bf{v}}, {\bf{h}})}-\underbrace{\sum_{i=1}^{K}c_{i,2}({\bf{v}})}_{C_{2}({\bf{v}},{\bf{h}})},
\end{eqnarray}
is a DC function, as both the functions $C_{1}({\bf{v}}, {\bf{h}})$
and $C_{2}({\bf{v}}, {\bf{h}})$ are convex in ${\bf{v}}$. Furthermore,
for any $z_1, z_2\in\mathbb{R}$ and $z=z_1-z_2$, we have $z^{+}=\max\{z_1,z_2\}-z_2$.
Therefore, 
\begin{eqnarray}
\label{bb1}
\psi\left(\max_{1\le k\le K}d_{k}({\bf{v}}),
\nu\right)={1\over{\nu}}[m({\bf{v}}, \nu)-m({\bf{v}}, 0)],
\end{eqnarray}
is a DC function of ${\bf{v}}$, as 
\begin{eqnarray}
\label{bb_2}
 m({\bf{v}}, \nu)=\max\{\nu+C_{1}({\bf{v}}, {\bf{h}}), C_{2}({\bf{v}},{\bf{h}})\},
\end{eqnarray}
is a convex function of ${\bf{v}}$. According to \cite[Proposition 2.1]{horst1999dc},
$\hat{f}({\bf{v}}, \nu)=\mathbb{E}[\psi\left(\max_{1\le k\le K}d_{k}({\bf{v}},{\bf{h}}_k),
\nu\right)]$ is a DC function on $\mathcal{V}$. Therefore, the proof is completed.

\section{Proof of Theorem {\ref{theorem_DCP}}}
\label{proof_DCP}
In order to prove Theorem {\ref{theorem_DCP}}, we need to prove
the following equality:
\begin{eqnarray}
\label{pt1}
\inf_{\nu>0} \hat{f}({\bf{v}}, \nu)=f({\bf{v}}). 
\end{eqnarray}
First, we need to prove the monotonicity of the function $\hat{f}({\bf{v}},
\nu)$ in the variable $\nu$. 
According to (\ref{bb1}) and (\ref{bb_2}), the function $\hat{f}({\bf{v}},
\nu)$ can be rewritten as
\begin{eqnarray}
\hat{f}({\bf{v}},\nu)={\mathbb{E}}[\pi(\nu, C_{1}({\bf{v}}, {\bf{h}}),
C_2({\bf{v}}, {\bf{h}}))],
\end{eqnarray}
where 
\begin{eqnarray}
\pi(\nu, z_1,z_2)\triangleq {1\over{\nu}}\left[\max\{\nu+z_1,z_2\}-\max\{z_1,z_2\}\right],
\end{eqnarray}
for any $z_1,z_2\in\mathbb{R}$ and $\nu>0$. Therefore, we only need to prove
the monotonicity of the function $\pi(\nu, z_1, z_2)$ in the variable $\nu$.

Define $z\triangleq z_1-z_2$, then we have
\begin{eqnarray}
\pi(\nu, z_1,z_2)=\left(1+{1\over{\nu}} z\right)1_{(-\nu,0]}(z)+1_{(0,+\infty)}(z).
\end{eqnarray} 
For any $\nu_1>\nu_2>0$ and any $z_1, z_2\in\mathbb{R}$, we have 
\begin{eqnarray}
&&\pi(\nu_1, z_1,z_2)-\pi(\nu_2, z_1,z_2)=\left(1+{1\over{\nu_1}} z\right)1_{(-\nu_1,-\nu_2]}(z)+\nonumber\\
&&z\left({1\over{\nu_1}}-{1\over{\nu_2}}\right)1_{(-\nu_2,0)}(z)\ge 0.
\end{eqnarray}
Therefore, $\hat{f}({{\bf{v}}, \nu})$ is nondecreasing in $\nu$ for $\nu>0$.
Hence, we have
\begin{eqnarray}
\label{nonde}
\!\!\inf_{\nu>0} \hat{f}({\bf{v}}, \nu)=\lim _{\nu\searrow 0}\hat
f({\bf{v}}, \nu)=\lim_{\nu\searrow 0} {1\over{\nu}}[u({\bf{v}},
\nu)-u ({\bf{v}}, 0)], 
\end{eqnarray}
where $\nu\searrow 0$ indicates that $\nu$ decreasingly goes to $0$. {{Thus, based
on (\ref{nonde}), in order to prove (\ref{pt1}), we only
need to prove
\begin{eqnarray}
\lim_{\nu\searrow 0} {1\over{\nu}}[u({\bf{v}}, \nu)-u ({\bf{v}}, 0)]=f({\bf{v}}).
\end{eqnarray}
Furthermore, if the partial derivation of $u({\bf{v}},\nu)$
exists, we have 
\begin{eqnarray}
\lim_{\nu\searrow 0} {1\over{\nu}}[u({\bf{v}}, \nu)-u ({\bf{v}}, 0)]={\partial\over{\partial
\nu}}u({\bf{v}},0).
\end{eqnarray}
Therefore, we need to prove that ${\partial\over{\partial \nu}}u({\bf{v}},\nu)$
exists and ${\partial\over{\partial \nu}}u({\bf{v}},0)=f({\bf{v}})$.
}}

According to (\ref{bb_2}), we have $u({\bf{v}}, \nu)={\mathbb{E}}[m({\bf{v}},
{\bf{h}}, \nu)]=\mathbb{E}[\max\{\nu+C_1({\bf{v}},{\bf{h}}), C_{2}({\bf{v}},{\bf{h}})\}]$. As
\begin{eqnarray}
{\partial\over{\partial\nu}}\left(\max\{\nu+z_1,z_2\}\right)=1_{(-\nu,+\infty)}(z),
\end{eqnarray}
for any $z\ne-\nu$, and 
${\rm{Pr}}\left\{\max_{1\le k\le K} d_{k}({\bf{v}}, {\bf{h}})=-\nu\right\}=0$,
where $\max\nolimits_{1\le k\le K} d_{k}({\bf{v}}, {\bf{h}})\triangleq
C_{1}({\bf{v}},
{\bf{h}})-C_{2}({\bf{v}},
{\bf{h}})$ (\ref{bb3}), we conclude that ${\partial\over{\partial\nu}}u({\bf{v}},\nu)$
exists.

Let $\mathcal{T}\triangleq(-T, T)$ with $T>0$ being an open set such that
the cumulative distribution
function $ F({\bf{v}}, \nu)\triangleq{\rm{Pr}}\{\max_{1\le k\le K} d_{k}({\bf{v}})\le
\nu)\}$ of the random variable $\left(\max_{1\le k\le K} d_{k}({\bf{v}})\right)$ is continuously
differentiable for any $\nu\in\mathcal{T}$. Next we will show that 
\begin{eqnarray}
\label{ddr}
\!\!\!\!\!\!\!{\partial\over{\partial \nu}}u({\bf{v}},\nu)&=&\lim_{\delta\rightarrow
0}{1\over{\delta}}{\mathbb{E}}[{m({\bf{v}}, {\bf{h}}, \nu+\delta)-m({\bf{v}},{\bf{h}}, \nu)}]\nonumber\\
&=&{\rm{Pr}}\left\{\max_{1\le k\le K} d_{k}({\bf{v}})>-\nu\right\}=1\!-\!F({\bf{v}},
-\nu).
\end{eqnarray}

For any $\nu\in\mathcal{T}$ and ${\bf{v}}\in\mathcal{V}$, define the random
variable $X(\delta)\triangleq[{m({\bf{v}},
{\bf{h}}, \nu+\delta)-m({\bf{v}},{\bf{h}},\nu)}]/\delta$, then
we have the following two facts: 
\begin{enumerate}
\item The limit of $X(\delta)$ exists and we have
\begin{eqnarray}
\lim_{\delta\rightarrow 0} X(\delta)=1_{(-\nu, +\infty)}\left(\max\limits_{1\le k\le
K} d_{k}({\bf{v}}, {\bf{h}})\right),  
\end{eqnarray}
with probability one. 
\item $X(\delta)$ is dominated by a constant $C>0$, i.e., 
$|X(\delta)|\le C$,
where $0<C<\infty$. This can be justified by
\begin{eqnarray}
|X(\delta)|&=&{1\over{h}}|{m({\bf{v}},
{\bf{h}}, \nu+\delta)-m({\bf{v}}, {\bf{h}},\nu)}|\nonumber\\
&=& {1\over{\delta}}|[\delta+Q({\bf{v}}, {\bf{h}}, \nu)]^{+}-[Q({\bf{v}}, {\bf{h}}, \nu)]^{+}|\le 1,\nonumber
\end{eqnarray} 
where  $Q({\bf{v}}, {\bf{h}}, \nu)\triangleq\nu+\max\nolimits_{1\le
k\le K} d_{k}({\bf{v}}, {\bf{h}})$ and the last inequality is based on the fact $|[x]^{+}-[y]^{+}|\le |x-y|$.
\end{enumerate}
From the above two facts on the random variable $X(\delta)$, by the
dominated convergence theorem to interchange an expectation
and the limit as $\delta\rightarrow 0$, and together with \cite[Proposition 1]{broadie1996estimating},
we have 
\begin{eqnarray}
\label{a3}
{\partial\over{\partial \nu}}u({\bf{v}},\nu)&=&\lim _{\delta\rightarrow
0} \mathbb{E}[X(\delta)]=\mathbb{E}[\lim\nolimits_{\delta\rightarrow
0} X(\delta)]\nonumber\\
&=&{\mathbb{E}} [1_{(-\nu, +\infty)}(\max\nolimits_{1\le k\le K} d_{k}({\bf{v}},
{\bf{h}}))]\nonumber\\
&=&1-F({\bf{v}},
-\nu).
\end{eqnarray}
Therefore, we complete the proof by
\begin{eqnarray}
\label{inff}
\inf_{\nu>0} \hat{f}({\bf{v}}, \nu)&=&\lim_{\nu\searrow 0} {1\over{\nu}}[u({\bf{v}},
\nu)-u ({\bf{v}}, 0)]\nonumber\\
&=&{\partial\over{\partial \nu}}u({\bf{v}},0)=1-F({\bf{v}},
0)=f({\bf{v}}).
\end{eqnarray}

\section{Proof of Lemma {\ref{lemma_GRAD}}}
\label{proof_GRAD}
It is well known that non-constant real-valued functions of complex variables
are not holomorphic (or $\mathbb{C}$-differentiable) \cite{hjorungnes2011complex}.
Thus, the real-valued
functions $d_{k}({\bf{v}}, {\bf{h}}_k)$ in (\ref{d_DC}) are not
differentiable
in the complex domain $\mathbb{C}^{NK}$ (i.e., with respect to the complex
vector ${\bf{v}}$).
Define a real-valued function 
$m({\bf{v}}, {\bf{h}}, \nu)\triangleq\max_{1\le k\le K+1} s_{k}({\bf{v}},
{\bf{h}}, \nu)$,
which is convex in ${\bf{v}}$. Although this function is not holomorphic
in ${\bf{v}}$, it can be viewed
as a function of both ${\bf{v}}$ and its complex conjugate ${\bf{v}}^{*}$,
i.e., $m({\bf{v}}, {\bf{v}}^{*},{\bf{h}}, \nu)$. It is easy to
verify that the function $m({\bf{v}}, {\bf{v}}^{*},{\bf{h}}, \nu)$
is holomorphic in ${\bf{v}}$ for a fixed ${\bf{v}}^{*}$ and is also holomorphic
in ${\bf{v}}^{*}$ for a fixed ${\bf{v}}$. Proving Lemma {\ref{lemma_GRAD}}
is equivalent to proving that the gradient of ${\mathbb{E}}[m({\bf{v}},
{\bf{h}}, \nu)]$ with respect to ${\bf{v}}^{*}$ exists and equals
\begin{eqnarray}
\label{finalp1}
\nabla_{{\bf{v}}^{*}}{\mathbb{E}}[m({\bf{v}}, {\bf{h}},\nu)]={\mathbb{E}}[\nabla_{{\bf{v}}^*}m({\bf{v}},{\bf{h}}, \nu)].
\end{eqnarray}

Based on
the chain rule \cite{hjorungnes2011complex}, the complex gradient of the
function $m({\bf{v}},{\bf{h}}, \nu)$ with respect to ${\bf{v}}^{*}$
exists and is given by
\begin{eqnarray}
\nabla_{{\bf{v}}^{*}}m({\bf{v}},{\bf{h}}, \nu)\triangleq{{\partial{m({\bf{v}},{\bf{h}}, \nu)}}\over{\partial{\bf{v}}^{*}}}={{\partial
s_{k^{\star}}({\bf{v}},{\bf{h}}, \nu)}\over{\partial {\bf{v}}^{*}}},
\end{eqnarray} 
with probability one, where $k^{\star}=\arg\max\limits_{1\le k\le K+1} s_{k}({\bf{v}},{\bf{h}}, \nu)$. It is a vector operator and gives the direction
of the steepest ascent of a real scalar-valued function.

Denote ${{\partial{m({\bf{v}},{\bf{h}}, \nu)}}\over{\partial{\bf{v}}^{*}}}\triangleq[{{\partial{m}}\over{\partial{{v}}_{i}^{*}}}]_{1\le
i\le NK}$ and ${{\partial{s_{k^{\star}}({\bf{v}}, {\bf{h}}_k, \nu)}}\over{\partial{\bf{v}}^{*}}}\triangleq[{{\partial{s_{k^{\star}}}}\over{\partial{{v}}_{i}^{*}}}]_{1\le
i\le NK}$, where ${\bf{v}}=[v_1,v_2,\dots, v_{NK}]$, and define the following
complex random variable
\begin{eqnarray}
Y(\Delta v_{i}^{*})\triangleq{1\over{{\Delta v_{i}^{*}}}}{[m({\bf{v}}_{-i},
v_{i}^{*}+\Delta v_{i}^{*})-m({\bf{v}}_{-i}, {{v}}_i^{*})]},
\end{eqnarray}
where ${\bf{v}}_{-i}\triangleq[v_{k}]_{k\ne i}$, $\Delta v_{i}^{*}\in\mathbb{C}$
and $m({\bf{v}})\triangleq m({\bf{v}},{\bf{h}}, \nu)$ for simplicity,
then we have the following two facts on the random variable $Y(\Delta
v_{i}^{*})$:
\begin{enumerate}
\item The limit of $Y(\Delta v_{i}^{*})$ exists and equals
\begin{eqnarray}
\label{fact1}
\lim_{\Delta v_{i}^*\rightarrow 0} Y(\Delta v_{i}^{*})= {{\partial{s_{
k^{\star}}}}\over{\partial{{v}}_{i}^{*}}},
\end{eqnarray}
with probability one.
\item The random variable is dominated by a random variable $Z$ with $\mathbb{E}[Z]\le
+\infty$, i.e.,
\begin{eqnarray}
\label{fact2}
|Y(\Delta {{v}}_{i}^{*})|\le Z, \forall i,
\end{eqnarray}
which can be verified
by the following lemma.  
\end{enumerate}

\begin{lemma}
For any ${\bf{x}}, {\bf{y}}\in\mathcal{V}$, there exists a random variable
$Z$ with
$\mathbb{E}[Z]\le \infty$ such that
\begin{eqnarray}
\label{lip}
|m({\bf{x}},{\bf{h}}, \nu)-m({\bf{y}},{\bf{h}}, \nu)|\le
Z\|{\bf{x}}-{\bf{y}}\|.
\end{eqnarray}

\begin{IEEEproof}
As $m({\bf{v}})$ is convex in ${\bf{v}}$, we have\begin{eqnarray}
m({\bf{x}})&\ge& m({\bf{y}})+2\langle\nabla_{{\bf{v}}^{*}}m({\bf{y}}),
{\bf{x}}-{\bf{y}}\rangle,\\
m({\bf{y}})&\ge& m({\bf{x}})+2\langle\nabla_{{\bf{v}}^{*}}m({\bf{x}}),
{\bf{y}}-{\bf{x}}\rangle.
\end{eqnarray}
Based on the above two inequalities and by the Cauchy-Schwarz inequality,
we have
\begin{eqnarray}
\label{lp_1}
|m({\bf{x}})-m({\bf{y}})|\le2\left(\max_{{\bf{v}}={\bf{x}},
{\bf{y}}} \|\nabla_{{\bf{v}}^{*}}m({\bf{v}})\|\right)\|{\bf{x}}-{\bf{y}}\|.
\end{eqnarray}
Furthermore, for $1\le k\le K$, we have 
\begin{eqnarray}
\|\nabla_{{\bf{v}}^{*}}s_{k}({\bf{v}})\|&=&\left({\sum_{i\ne k}\left\|\left({\bf{h}}_{k}{\bf{h}}_{k}^{\sf{H}}+{1\over{\gamma_{k}^2}}{\bf{h}}_i{\bf{h}}_{i}^{\sf{H}}\right){\bf{v}}_{i}\right\|^2}\right)^{1/2}\nonumber\\
&\le&\max_{i\ne k}\|{\bf{v}}_i\| \left({\sum_{i\ne k}\left\|\left({\bf{h}}_{k}{\bf{h}}_{k}^{\sf{H}}+{1\over{\gamma_{k}^2}}{\bf{h}}_i{\bf{h}}_{i}^{\sf{H}}\right)\right\|^2}\right)^{1/2}\nonumber\\
&=& Z_1,
\end{eqnarray}
where $Z_{1}$ is a random variable with $\mathbb{E}[Z_1]\le+\infty$, and
for
$k=K+1$, we have
\begin{eqnarray}
\|\nabla_{{\bf{v}}^{*}}s_{K+1}({\bf{v}})\|&=&\left({\sum_{i=1}^{K}\left\|{1\over{\gamma_{i}^2}}{\bf{h}}_{i}{\bf{h}}_{i}^{\sf{H}}{\bf{v}}_{i}\right\|^2}\right)^{1/2}\nonumber\\
&\le&\max_{1\le i\le K}\|{\bf{v}}_{i}\|\left({\sum_{i=1}^{K}\left\|{1\over{\gamma_{i}^2}}{\bf{h}}_{i}{\bf{h}}_{i}^{\sf{H}}\right\|^2}\right)^{1/2}\nonumber\\
&=&Z_2,
\end{eqnarray}
where $Z_{2}$ is a random variable with $\mathbb{E}[Z_2]\le+\infty$. Therefore,
letting $Z\triangleq\max\{Z_1, Z_2\}$ with $\mathbb{E}[Z]<+\infty$, we have
\begin{eqnarray}
\label{lp_2}
\nabla_{{\bf{v}}^{*}}m({\bf{v}})={{\partial
s_{k^{\star}}({\bf{v}})}\over{\partial {\bf{v}}^{*}}}\le\max\{Z_1, Z_2\}=Z.
\end{eqnarray}
According to (\ref{lp_1}) and (\ref{lp_2}), we have the inequality (\ref{lip}).
\end{IEEEproof}
\end{lemma}

Based on the above two facts (\ref{fact1}) and (\ref{fact2}) on the random
variable $Y(\Delta
v_{i}^{*})$, and by the
dominated convergence theorem to interchange an expectation
and the limit as $\Delta v_{i}^{*}\rightarrow 0$ and \cite[Proposition 1]{broadie1996estimating},
we have 
\begin{eqnarray}
\!\!\!\!\lim _{\Delta
v_{i}^{*}\rightarrow
0} \mathbb{E}[Y(\Delta
v_{i}^{*})]=\mathbb{E}\left[\lim _{\Delta
v_{i}^{*}\rightarrow
0} Y(\Delta
v_{i}^{*})\right]=\mathbb{E}\left[{{\partial{s_{k^{\star}}}}\over{\partial{{v}}_{i}^{*}}}\right].
\end{eqnarray}
Based on the fact
\begin{eqnarray}
\nabla_{{\bf{v}}^{*}}{\mathbb{E}}[m({\bf{v}}, {\bf{h}}, \nu)]=\left[\lim
_{\Delta
v_{i}^{*}\rightarrow
0} \mathbb{E}[Y(\Delta
v_{i}^{*})]\right]_{1\le i\le NK},
\end{eqnarray}
we get (\ref{finalp1}) and thus complete the proof.

\section{Proof of Theorem {\ref{theorem_CON}}}
\label{proof_CON}
For simplicity, we only consider the case with real variables and
functions. The extension to complex variables is straightforward. Specifically, define $\mathcal{D}_{0}=\{{\bf{v}}\in\mathcal{V}: f({\bf{v}})\le\epsilon\}$ as the feasible set of the SCB problem $\mathscr{P}_{\textrm{SCB}}$. To ensure the existence of the KKT paris for the SCB problem $\mathscr{P}_{\textrm{SCB}}$, we assume the following constraint qualification \cite[Corollary 6.15]{rockafellar1998variational} for program $\mathscr{P}_{\textrm{SCB}}$, i.e., for any feasible point ${\bf{v}}\in\mathcal{D}_{0}$, $\lambda=0$
is the only value that satisfies the following linear system:
\begin{eqnarray}
\label{cq}
-\lambda\nabla_{\bf{v}}f({\bf{v}})\in\mathcal{N}_{\mathcal{V}}({\bf{v}}), \lambda [f({\bf{v}})-\epsilon]=0
\end{eqnarray}
where $\lambda\ge 0$, and ${\mathcal{N}}_{\mathcal{V}}({\bf{v}})$ is the normal cone to the convex set $\mathcal{V}$ at ${\bf{v}}$, i.e.,
\begin{eqnarray}
\mathcal{N}_{\mathcal{V}}({\bf{v}})=\{{\bf{x}}|\langle{\bf{x}}, {\bf{y}}-{\bf{v}}\rangle\le 0, \forall {\bf{y}}\in\mathcal{V}\}.
\end{eqnarray}With this constraint qualification, we have the KKT pairs (${\bf{v}}^{\star}, {\lambda}^{\star}$) \cite[Corollary 6.15]{rockafellar1998variational}
for the SCB problem as
\begin{eqnarray}
\Omega_{0}:
\left\{\begin{array}{l}
-[\nabla_{\bf{v}} f_{0}({\bf{v}}^{\star})+\lambda^{\star}\nabla_{\bf{v}}f({\bf{v}}^{\star})]\in \mathcal{N}_{\mathcal{V}}({\bf{v}}^{\star})\\
\lambda^{\star}[f({\bf{v}}^{\star})-\epsilon]=0\\
\lambda^{\star}\ge0, {\bf{v}}^{\star}\in\mathcal{V},
\end{array}\right.
\end{eqnarray}
where $f_{0}({\bf{v}})=\|{\bf{v}}\|^2$ is the objective function of $\mathscr{P}_{\textrm{SCB}}$. 

Similarly, let $({\bf{v}}^{\star}, \kappa^{\star}, \lambda^{\star})$ be a KKT pair of the joint approximation program $\tilde{\mathscr{P}}_{\textrm{DC}}$ as follows
\begin{eqnarray}
\Omega:
\left\{\begin{array}{l}
-\left\{\nabla_{\bf{v}} f_{0}({\bf{v}}^{\star})+
\lambda^{\star}\nabla_{\bf{v}}[u({\bf{v}}^{\star},\kappa^{\star})-\kappa^{\star}\epsilon-u({\bf{v}}^{\star},0)]\right\}\\\
\in
\mathcal{N}_{\mathcal{V}}({\bf{v}}^{\star}),\\
-\left\{\lambda^{\star}\nabla_{\kappa}[u({\bf{v}}^{\star},\kappa^{\star})-\kappa^{\star}\epsilon-u({\bf{v}}^{\star},0)]\right\}\in \mathcal{N}_{(0,+\infty)}(\kappa^{\star}),\\
\lambda^{\star}[u({\bf{v}}^{\star},\kappa^{\star})-\kappa^{\star}\epsilon-u({\bf{v}}^{\star},0)]=0\nonumber\\
\lambda^{\star}\ge0, {\bf{v}}^{\star}\in\mathcal{V}, \kappa >0.
\end{array}\right.
\end{eqnarray}

In order to prove Theorem {\ref{theorem_CON}}, we first prove the following lemma illustrating the relationship between $\Omega_{0}$ and $\Omega$.

\begin{lemma}
\label{kktopt}
Suppose that there exists $({\bf{v}}^{[j]}, {\kappa}^{[j]}, \lambda^{[j]}) \in \Omega$, such that $({\bf{v}}^{[j]}, \kappa^{[j]}, \lambda^{[j]})\rightarrow (\hat{\bf{v}}, 0, \hat{\lambda})$, then we have that $(\hat{\bf{v}}, \hat{\lambda})\in \Omega_{0}$.
\begin{IEEEproof}
We only need to consider two cases in terms of $\lambda^{[j]}$ being zeros or not. 

{{Case one}}: suppose there exists a subsequence $\{\lambda^{[k_i]}\}$ of $\{\lambda^{[j]}\}$ such that $\lambda^{[k_i]}=0, i=0,1,2,\dots$.
As $\lambda^{[k_i]}$'s belong to $\Omega$, we have $-\nabla_{\bf{v}} f_{0}({\bf{v}}^{[k_i]})\in \mathcal{N}_{\mathcal{V}}({\bf{v}}^{[k_i]})$,
which implies that $-\nabla_{\bf{v}} f_{0}(\hat{\bf{v}})\in
\mathcal{N}_{\mathcal{V}}(\hat{\bf{v}})$,
as $i\rightarrow \infty$. This indicates $(\hat{\bf{v}}, 0)\in \Omega_0$.

{{Case two}}: suppose that $\lambda^{[n]}\ne 0$, for sufficiently large $n$. In this case, we have $\nabla_{\kappa}[u({\bf{v}}^{[n]},\kappa^{[n]})-\kappa^{[n]}\epsilon]=0$,
as $\kappa^{[n]}>0$ and $\mathcal{N}_{(0,+\infty)}(\kappa^{[n]})=0$. Based on (\ref{inff}),  let $n\rightarrow \infty$ such that $\kappa^{[n]}\rightarrow 0$, we have
\begin{eqnarray}
\label{kkt4}
f(\hat{\bf{v}})-\epsilon=0.
\end{eqnarray}

Furthermore, as $\kappa^{[n]}\ne 0$, based on the KKT pairs in $\Omega$, we have
\begin{eqnarray}
\label{kkt1}
&&-\nabla_{\bf{v}} f_{0}({\bf{v}}^{[n]})-\lambda^{[n]}\kappa^{[n]}\left\{{{\nabla_{\bf{v}}[u({\bf{v}}^{[n]},\kappa^{[n]})-u({\bf{v}}^{[n]},0)]}\over{\kappa^{[n]}}}\right\}\nonumber\\
&&\in\mathcal{N}_{\mathcal{V}}({\bf{v}}^{[n]}),
\end{eqnarray}
and
\begin{eqnarray}
\label{kkt2}
\lambda^{[n]}\kappa^{[n]}\left\{\nabla_{\kappa}[u({\bf{v}}^{[n]},\kappa^{[n]})-\kappa^{[n]}\epsilon]\right\}=0.
\end{eqnarray}

According
to (\ref{a3}), we have
\begin{eqnarray}
{{\partial}\over{\partial{\kappa}}}\nabla_{{\bf{v}}}u({\bf{v}}^{[n]},
\kappa^{[n]})&=&\nabla_{{\bf{v}}}\left({{\partial}\over{\partial{\kappa}}}u({\bf{v}}^{[n]},
\kappa^{[n]})\right)\nonumber\\
&=&-\nabla_{{\bf{v}}}F({\bf{v}}^{[k]}, -\kappa^{[n]}).
\end{eqnarray} 
Therefore, we have
\begin{eqnarray}
\label{kkt3}
\!\!\!\!\!\!\!\!\!\!\!\!\!\!\!\!\!\!\!\!\!&&\lim\limits_{n\rightarrow+\infty}{{\nabla_{{\bf{v}}}u({\bf{v}}^{[n]},
\kappa^{[n]})-\nabla_{{\bf{v}}}u({\bf{v}}^{[k]},
0)}\over{\kappa^{[n]}}}=\nonumber\\
\!\!\!\!\!\!\!\!&&-\lim\limits_{n\rightarrow+\infty}\nabla_{{\bf{v}}}F({\bf{v}}^{[n]},
-\bar\kappa^{[n]})=-\nabla_{{\bf{v}}}F(\hat{\bf{v}}, 0)=\nabla_{{\bf{v}}}f(\hat{\bf{v}}),
\end{eqnarray}
where $\bar{\kappa}^{[n]}\in(0,\kappa^{[n]}), \forall n$, due to the mean-value
theorem.

Dividing both sides of equations (\ref{kkt1}) and (\ref{kkt2}) by $\lambda^{[n]}\kappa^{[n]}$, respectively, 
let $n\rightarrow \infty$ and  suppose that $\lambda^{[n]}\kappa^{[n]}\rightarrow +\infty$, based on (\ref{kkt4}) and (\ref{kkt3}), we have
\begin{eqnarray}
-\nabla_{\bf{v}}f(\hat{\bf{v}})\in\mathcal{N}_{\mathcal{V}}(\hat{\bf{v}}), f(\hat{\bf{v}})-\epsilon=0.
\end{eqnarray}
However, this contradicts the constraint qualification (\ref{cq}). Therefore, we conclude that $\lambda^{[n]}\kappa^{[n]}\nrightarrow +\infty$. We thus assume that $\lambda^{[n]}\kappa^{[n]}\rightarrow \hat{\lambda}$ with $0\le \hat{\lambda}<+\infty$. Let $n\rightarrow\infty$, based on (\ref{kkt4}), (\ref{kkt1}), (\ref{kkt2}) and (\ref{kkt3}), we obtain
\begin{eqnarray}
\!\!\!-\left\{\nabla_{\bf{v}} f_{0}(\hat{\bf{v}})+\hat\lambda\nabla_{\bf{v}}f(\hat{\bf{v}})\right\}\in
\mathcal{N}_{\mathcal{V}}(\hat{\bf{v}}), \hat{\lambda}[f(\hat{\bf{v}})-\epsilon]=0.
\end{eqnarray}
This indicates that $(\hat{\bf{v}}, \hat{\lambda})\in\Omega_{0}$. We thus complete the proof. 
\end{IEEEproof}
\end{lemma}

Based on  Lemma \ref{kktopt}, we further investigate whether $\kappa$ converges to zero. The answer is positive in most scenarios except two special cases. Suppose that $(\hat{\bf{v}}, \hat{\kappa})$ is a KKT point of the problem $\tilde{\mathscr{P}}_{\textrm{DC}}$. We consider two particular cases in terms of whether the SCB program $\mathscr{P}_{\textrm{SCB}}$ attaining its optimal value at the interior point or not.

Case one: When the SCB program $\mathscr{P}_{\textrm{SCB}}$ attains the optimal value at the interior point of its feasible region, then program $\tilde{\mathscr{P}}_{\textrm{SCB}}$ also attains its optimal value at the interior point of its feasible region based on Theorem {\ref{dcapp}}. In this scenario, the DC constraint in $\tilde{\mathscr{P}}_{\textrm{SCB}}$ does not need to be tight. Thus, $\hat{\kappa}$ is not necessary to be zero and it has multiple choices, while $(\hat{\bf{v}}, 0)$ still belongs to $\Omega_{0}$.

Case two: When all the optimal solutions of the SCB program ${\mathscr{P}}_{\textrm{SCB}}$ make the probability constraint tight. In this scenario, we have  
$[u(\hat{\bf{v}}, \hat\kappa)-\hat\kappa\epsilon]-u(\hat{\bf{v}},
0)=0$. This reveals that $\kappa=0$ is a minimizer of the function $[u(\hat{\bf{v}},  {\kappa})-\kappa\epsilon]$
with respect to $\kappa$,
i.e.,
\begin{eqnarray}
\label{z2}
{\rm{Pr}}\left\{\max_{1\le
k\le K}d_{k}(\hat{\bf{v}}, {\bf{h}})>0\right\}= \epsilon,
\end{eqnarray}
where the calculation is based on (\ref{ddr}). On the other hand, as
$\hat\kappa$ satisfies the KKT conditions of program $\tilde{\mathscr{P}}_{\textrm{DC}}$, we have 
\begin{eqnarray}
\label{dd1}
\nabla_{\kappa}[u(\hat{\bf{v}},  \hat{\kappa})-\hat{\kappa}\epsilon]=0.
\end{eqnarray}
According to \cite[Theorem 10]{Rockafellar2002conditional}
and \cite[Appendix A4]{Hong_2013smooth}, the minimizer (i.e., $\hat\kappa\ne
0$ in (\ref{dd1})) of the function $[u(\hat{\bf{v}},  {\kappa})-\kappa\epsilon]$ with
respect to $\kappa$ satisfies \begin{eqnarray}
\label{z1}
{\rm{Pr}}\left\{\max_{1\le
k\le K}d_{k}({\bf{v}}, {\bf{h}})>-\hat\kappa\right\}\le \epsilon.
\end{eqnarray}
Combining (\ref{z2}) and (\ref{z1}), we conclude that  ${\rm{Pr}}\left\{\max_{1\le
k\le K}d_{k}(\hat{\bf{v}}, {\bf{h}})\in(-\hat\kappa, 0]\right\}= 0$.
This implies that the optimization variable $\kappa$ in $\tilde{\mathscr{P}}_{\textrm{DC}}$ converges
to zero, if
for any $c>0$, we have
\begin{eqnarray}
\label{zzc}
{\rm{Pr}}\left\{\max_{1\le k\le K}d_{k}(\hat{\bf{v}}, {\bf{h}})\in[-c,
0]\right\}\ne 0.
\end{eqnarray}
From numerical examples in Section {\ref{sim}}, we will demonstrate that
variable $\kappa$ will  indeed converge to zero. 

Finally, based on Lemma {\ref{kktopt}}, we only need to prove that the sequence generated by the stochastic DC programming algorithm converges to a KKT point of the program $\tilde{\mathscr{P}}_{\textrm{DC}}$. This directly follows  \cite[Property 3]{hong2011sequential}. We thus complete the proof.

\section{Proof of Theorem {\ref{theorem_est}}}
\label{proof_est}
By \cite[Theorem 7.50]{shapiro2009lectures} and \cite[Theorem 6]{hong2011sequential},
we have that the SAA estimate $\bar{l}({\bf{v}},\kappa; {\bf{v}}^{[j]}, \kappa^{[j]})$ (\ref{linear_est}) converges
to $l({\bf{v}}, \kappa; {\bf{v}}^{[j]}, \kappa^{[j]})$ uniformly on the convex compact set $\mathcal{V}$
with probability one as $M\rightarrow +\infty$, i.e.,
\begin{eqnarray}
\!\!\!\!\!\!\!\sup_{{\bf{v}}\in\mathcal{V}}|\bar{l}({\bf{v}}, \kappa; {\bf{v}}^{[j]}, \kappa^{[j]})-l({\bf{v}}, \kappa;
{\bf{v}}^{[j]}, \kappa^{[j]})|\rightarrow 0, M\!\!\rightarrow \!+\infty,
\end{eqnarray} 
with probability one. Furthermore, by \cite[Theorem 5.5]{shapiro2009lectures}, we have $V_{M}^{\star}({\bf{v}}^{[j]}, \kappa^{[j]})\rightarrow V^{\star}({\bf{v}}^{[j]}, \kappa^{[j]})$
and $\mathbb{D}(\mathcal{P}^{\star}_{M}({\bf{v}}^{[j]}, \kappa^{[j]}), \mathcal{P}^{\star}({\bf{v}}^{[j]}, \kappa^{[j]}))\rightarrow
0$ with probability one as $M\rightarrow +\infty$. Therefore, we complete
the proof.

\bibliographystyle{ieeetr}
\bibliography{/Users/yuanmingshi/Dropbox/Reference/Reference}

\begin{thebibliography}{10}

\bibitem{Gesbert_JSAC10}
D.~Gesbert, S.~Hanly, H.~Huang, S.~Shamai~Shitz, O.~Simeone, and W.~Yu,
  ``Multi-cell {MIMO} cooperative networks: A new look at interference,'' {\em
  IEEE J. Sel. Areas Commun.}, vol.~28, pp.~1380--1408, Sep. 2010.

\bibitem{Foschini_2006network}
M.~K. Karakayali, G.~J. Foschini, and R.~A. Valenzuela, ``Network coordination
  for spectrally efficient communications in cellular systems,'' {\em IEEE
  Wireless Commun.}, vol.~13, pp.~56--61, Aug. 2006.

\bibitem{Jun_2009networked}
J.~Zhang, R.~Chen, J.~G. Andrews, A.~Ghosh, and R.~W. Heath, ``Networked {MIMO}
  with clustered linear precoding,'' {\em IEEE Trans. Wireless Commun.},
  vol.~8, pp.~1910--1921, Apr. 2009.

\bibitem{mobile2011c}
{China Mobile}, ``C-{RAN}: the road towards green {RAN},'' {\em White Paper,
  ver. 2.5}, Oct. 2011.

\bibitem{Yuanming_TWC2014}
Y.~Shi, J.~Zhang, and K.~B. Letaief, ``Group sparse beamforming for green
  {C}loud-{RAN},'' {\em IEEE Trans. Wireless Commun.}, vol.~13, pp.~2809--2823,
  May 2014.

\bibitem{Jindal_TC2010unified}
N.~Jindal and A.~Lozano, ``A unified treatment of optimum pilot overhead in
  multipath fading channels,'' {\em IEEE Trans. Commun.}, vol.~58,
  pp.~2939--2948, Oct. 2010.

\bibitem{love2008overview}
D.~J. Love, R.~W. Heath, V.~K. Lau, D.~Gesbert, B.~D. Rao, and M.~Andrews, ``An
  overview of limited feedback in wireless communication systems,'' {\em IEEE
  J. Sel. Areas Commun.}, vol.~26, pp.~1341--1365, Oct. 2008.

\bibitem{Tse_TIT2012completely}
M.~A. Maddah-Ali and D.~Tse, ``Completely stale transmitter channel state
  information is still very useful,'' {\em IEEE Trans. Inf. Theory}, vol.~58,
  pp.~4418--4431, Jul. 2012.

\bibitem{Jun_2009mode}
J.~Zhang, R.~W. Heath, M.~Kountouris, and J.~G. Andrews, ``Mode switching for
  the multi-antenna broadcast channel based on delay and channel
  quantization,'' {\em EURASIP Journal on Advances in Signal Processing},
  vol.~2009, Article ID 802548, 15 pages, 2009.

\bibitem{Yuaning_ICC2014}
Y.~Shi, J.~Zhang, and K.~Letaief, ``{CSI} overhead reduction with stochastic
  beamforming for cloud radio access networks,'' in {\em Proc. of IEEE Int.
  Conf. on Commun. (ICC), Sydney, Australia}, Jun. 2014.

\bibitem{Luo_arXiv2013}
M.~Razaviyayn, M.~Sanjabi, and Z.-Q. Luo, ``A stochastic successive
  minimization method for nonsmooth nonconvex optimization with applications to
  transceiver design in wireless communication networks,'' {\em CoRR},
  vol.~abs/1307.4457, 2013.

\bibitem{WeiYu_WC10}
H.~Dahrouj and W.~Yu, ``Coordinated beamforming for the multicell multi-antenna
  wireless system,'' {\em IEEE Trans. Wireless Commun.}, vol.~9,
  pp.~1748--1759, Sep. 2010.

\bibitem{ben2009robust}
A.~Ben-Tal, L.~El~Ghaoui, and A.~Nemirovski, {\em Robust optimization}.
\newblock Princeton University Press, 2009.

\bibitem{shapiro2009lectures}
A.~Shapiro, D.~Dentcheva, and A.~P. Ruszczy{\'n}ski, {\em Lectures on
  stochastic programming: modeling and theory}, vol.~9.
\newblock SIAM, 2009.

\bibitem{Emil_TSP2012}
E.~Bj{\"o}rnson, G.~Zheng, M.~Bengtsson, and B.~Ottersten, ``Robust monotonic
  optimization framework for multicell {MISO} systems,'' {\em IEEE Trans.
  Signal Process.}, vol.~60, pp.~2508--2523, May 2012.

\bibitem{bertsimas2011theory}
D.~Bertsimas, D.~B. Brown, and C.~Caramanis, ``Theory and applications of
  robust optimization,'' {\em SIAM review}, vol.~53, no.~3, pp.~464--501, 2011.

\bibitem{gershman2010convex}
A.~B. Gershman, N.~D. Sidiropoulos, S.~Shahbazpanahi, M.~Bengtsson, and
  B.~Ottersten, ``Convex optimization-based beamforming: From receive to
  transmit and network designs,'' {\em IEEE Signal Process. Mag}, vol.~27,
  no.~3, pp.~62--75, 2010.

\bibitem{Bjornson_TCIT2013}
E.~Bj{\"o}rnson and E.~Jorswieck, ``Optimal resource allocation in coordinated
  multi-cell systems,'' {\em Foundations and Trends in Communications and
  Information Theory}, vol.~9, pp.~113--381, Jan. 2013.

\bibitem{Angela_TSP10}
W.-L. Li, Y.~J. Zhang, A.-C. So, and M.~Z. Win, ``Slow adaptive {OFDMA} systems
  through chance constrained programming,'' {\em IEEE Trans. Signal Process.},
  vol.~58, pp.~3858--3869, Jul. 2010.

\bibitem{Angela_JSAC2013}
A.~M.-C. So and Y.~J.~A. Zhang, ``Distributionally robust slow adaptive {OFDMA}
  with soft {Q}o{S} via linear programming,'' {\em IEEE J. Sel. Areas Commun.},
  vol.~31, pp.~947--958, May 2013.

\bibitem{Vicent_TSP2013}
V.~Lau, F.~Zhang, and Y.~Cui, ``Low complexity delay-constrained beamforming
  for multi-user {MIMO} systems with imperfect {CSIT},'' {\em IEEE Trans.
  Signal Process.}, vol.~61, pp.~4090--4099, Aug. 2013.

\bibitem{Xiaodong_2014power}
W.~Xu, A.~Tajer, X.~Wang, and S.~Alshomrani, ``Power allocation in {MISO}
  interference channels with stochastic {CSIT},'' {\em IEEE Trans. Wireless
  Commun.}, vol.~13, pp.~1716--1727, Mar. 2014.

\bibitem{wang2011probabilistic}
K.-Y. Wang, T.-H. Chang, W.-K. Ma, A.-C. So, and C.-Y. Chi, ``Probabilistic
  {SINR} constrained robust transmit beamforming: A {B}ernstein-type inequality
  based conservative approach,'' in {\em Proc. of IEEE Int. Conf. Speech Signal
  Process. (ICASSP)}, pp.~3080--3083, May 2011.

\bibitem{hong2011sequential}
L.~J. Hong, Y.~Yang, and L.~Zhang, ``Sequential convex approximations to joint
  chance constrained programs: A {M}onte {C}arlo approach,'' {\em Oper. Res.},
  vol.~59, pp.~617--630, May-Jun. 2011.

\bibitem{Campi_SIAM2008}
M.~C. Campi and S.~Garatti, ``The exact feasibility of randomized solutions of
  uncertain convex programs,'' {\em SIAM J. Optim.}, vol.~19, no.~3,
  pp.~1211--1230, 2008.

\bibitem{nemirovski2006convex}
A.~Nemirovski and A.~Shapiro, ``Convex approximations of chance constrained
  programs,'' {\em SIAM J. Optim.}, vol.~17, no.~4, pp.~969--996, 2006.

\bibitem{Luo2013unified}
M.~Razaviyayn, M.~Hong, and Z.-Q. Luo, ``A unified convergence analysis of
  block successive minimization methods for nonsmooth optimization,'' {\em SIAM
  J. Optim.}, vol.~23, no.~2, pp.~1126--1153, 2013.

\bibitem{rockafellar2000optimization}
R.~T. Rockafellar and S.~Uryasev, ``Optimization of conditional
  value-at-risk,'' {\em Journal of risk}, vol.~2, pp.~21--42, 2000.

\bibitem{Jeffrey_2011mimo}
B.~Nosrat-Makouei, J.~G. Andrews, and R.~W. Heath, ``{MIMO} interference
  alignment over correlated channels with imperfect {CSI},'' {\em IEEE Trans.
  Signal Process.}, vol.~59, no.~6, pp.~2783--2794, 2011.

\bibitem{horst1999dc}
R.~Horst and N.~V. Thoai, ``{DC} programming: overview,'' {\em J. Optimiz.
  Theory Appl.}, vol.~103, pp.~1--43, Oct. 1999.

\bibitem{boyd2004convex}
S.~P. Boyd and L.~Vandenberghe, {\em Convex optimization}.
\newblock Cambridge university press, 2004.

\bibitem{nesterov1994interior}
Y.~Nesterov, A.~Nemirovskii, and Y.~Ye, {\em Interior-point polynomial
  algorithms in convex programming}, vol.~13.
\newblock SIAM, 1994.

\bibitem{boyd2011distributed}
S.~Boyd, N.~Parikh, E.~Chu, B.~Peleato, and J.~Eckstein, ``Distributed
  optimization and statistical learning via the alternating direction method of
  multipliers,'' {\em Foundations and Trends in Machine Learning}, vol.~3,
  pp.~1--122, Jul. 2011.

\bibitem{Ye_2010distributionally}
E.~Delage and Y.~Ye, ``Distributionally robust optimization under moment
  uncertainty with application to data-driven problems,'' {\em Oper. Res.},
  vol.~58, pp.~595--612, May-Jun. 2010.

\bibitem{broadie1996estimating}
M.~Broadie and P.~Glasserman, ``Estimating security price derivatives using
  simulation,'' {\em Management Sci.}, vol.~42, pp.~269--285, Feb. 1996.

\bibitem{hjorungnes2011complex}
A.~Hj{\o}rungnes, {\em Complex-valued matrix derivatives: with applications in
  signal processing and communications}.
\newblock Cambridge University Press, 2011.

\bibitem{rockafellar1998variational}
R.~T. Rockafellar and R.~J.-B. Wets, {\em Variational analysis}, vol.~317.
\newblock Springer, 1998.

\bibitem{Rockafellar2002conditional}
R.~T. Rockafellar and S.~Uryasev, ``Conditional value-at-risk for general loss
  distributions,'' {\em Journal of Banking and Finance}, vol.~26, no.~7,
  pp.~1443--1471, 2002.

\bibitem{Hong_2013smooth}
Z.~Hu, L.~J. Hong, and L.~Zhang, ``A smooth monte carlo approach to joint
  chance-constrained programs,'' {\em IIE Transactions}, vol.~45, no.~7,
  pp.~716--735, 2013.

\end{thebibliography}

\end{document}